\documentclass[journal]{IEEEtran}
\usepackage{etoolbox}
\makeatletter
\patchcmd{\@makecaption}
  {\scshape}
  {\normalsize}
  {}
  {}
\makeatletter
\patchcmd{\@makecaption}
  {\\}
  {\normalsize{:}\ }
  {}
  {}
\makeatother
\def\tablename{\normalsize{TABLE}}

\makeatletter
\def\endthebibliography{%
  \def\@noitemerr{\@latex@warning{Empty `thebibliography' environment}}%
  \endlist
}
\makeatother
\IEEEoverridecommandlockouts

\usepackage{cite}
\usepackage{amsmath,amssymb,amsfonts}

\usepackage{algorithm}
\usepackage{array}
\usepackage[caption=false,font=footnotesize,labelfont=sf,textfont=sf]{subfig}
\usepackage{fixltx2e}
\usepackage{stfloats}
\usepackage{url}
\usepackage{algpseudocode}
\usepackage{graphicx}
\usepackage{textcomp}
\usepackage{xcolor}
\usepackage{mathrsfs}
\usepackage{float}
\usepackage{bm}
\usepackage{enumerate}
\usepackage{url}

\def\BibTeX{{\rm B\kern-.05em{\sc i\kern-.025em b}\kern-.08em
    T\kern-.1667em\lower.7ex\hbox{E}\kern-.125emX}}

\begin{document}

\title{Integrated Sensing and Communications for Low-Altitude Economy: A Deep Reinforcement Learning Approach}

\author{Xiaowen~Ye, Yuyi~Mao,~\IEEEmembership{Senior~Member,~IEEE}, Xianghao~Yu,~\IEEEmembership{Senior~Member,~IEEE},
Shu~Sun,~\IEEEmembership{Member,~IEEE}, Liqun~Fu,~\IEEEmembership{Senior~Member,~IEEE}, and Jie~Xu,~\IEEEmembership{Fellow,~IEEE}
\thanks{Xiaowen~Ye and Xianghao Yu are with the Department of Electrical Engineering, City University of Hong Kong, Hong Kong, China. (E-mail: xiaowen.ye@cityu.edu.hk, alex.yu@cityu.edu.hk). (\textit{Corresponding author: Xianghao Yu.})}
\thanks{Yuyi Mao is with the Department of Electrical and Electronic Engineering, The Hong Kong Polytechnic University, Hong Kong, China. (E-mail: yuyi-eie.mao@polyu.edu.hk).}
\thanks{Shu Sun is with the Department of Electronic Engineering and the Cooperative Medianet Innovation Center, Shanghai Jiao Tong University, Shanghai 200240, China. (E-mail: shusun@sjtu.edu.cn).}
\thanks{Liqun~Fu is with the School of Informatics, Xiamen University, Xiamen, China. (E-mail: liqun@xmu.edu.cn).}
\thanks{Jie~Xu is with the School of Science and Engineering(SSE), the Shenzhen Future Network of Intelligence Institute (FNii-Shenzhen), and the Guangdong Provincial Key Laboratory of Future Networks of Intelligence, The Chinese University of Hong Kong, Shenzhen, China. (E-mail: xujie@cuhk.edu.cn).}
}

\renewcommand{\thefootnote}{\arabic{footnote}}
\maketitle

\begin{abstract}
This paper studies an integrated sensing and communications (ISAC) system for low-altitude economy (LAE), where a ground base station (GBS) provides communication and navigation services for authorized unmanned aerial vehicles (UAVs), while sensing the low-altitude airspace to monitor the unauthorized mobile target. The expected communication sum-rate over a given flight period is maximized by jointly optimizing the beamforming at the GBS and UAVs' trajectories, subject to the constraints on the average signal-to-noise ratio requirement for sensing, the flight mission and collision avoidance of UAVs, as well as the maximum transmit power at the GBS. Typically, this is a sequential decision-making problem with the given flight mission. Thus, we transform it to a specific Markov decision process (MDP) model called episode task. Based on this modeling, we propose a novel LAE-oriented ISAC scheme, referred to as Deep LAE-ISAC (DeepLSC), by leveraging the deep reinforcement learning (DRL) technique. In DeepLSC, a reward function and a new action selection policy termed constrained noise-exploration policy are judiciously designed to fulfill various constraints. To enable efficient learning in episode tasks, we develop a hierarchical experience replay mechanism, where the gist is to employ all experiences generated within each episode to jointly train the neural network. Besides, to enhance the convergence speed of DeepLSC, a symmetric experience augmentation mechanism, which simultaneously permutes the indexes of all variables to enrich available experience sets, is proposed. Simulation results demonstrate that compared with benchmarks, DeepLSC yields a higher sum-rate while meeting the preset constraints, achieves faster convergence, and is more robust against different settings.
\end{abstract}


\section{Introduction}\label{intro}
\IEEEPARstart{L}{ow}-altitude economy (LAE), composed of various low-altitude flight activities involving unmanned and manned aircraft, such as unmanned aerial vehicles (UAVs) and electric vertical take-off and landing (eVTOL), has gained widespread attention from academia and industry \cite{jiang20236g}. It is expected that LAE could support a wide variety of important applications, including but not limited to transportation, environmental monitoring, tourism, and agriculture \cite{LAESurvey}. However, the successful implementation of LAE requires the strictly safe operation (e.g., collision-free) of all aircraft, especially in the presence of massive aircraft and unauthorized targets \cite{mu2023uav}. Therefore, it is imperative to provide seamless communication and navigation for authorized aircraft, as well as ubiquitous surveillance of unauthorized targets in the low-altitude airspace.

One possible solution that could facilitate the implementation of LAE is to jointly adopt wireless communications and radar sensing technologies. Conventionally, these two technologies are designed and implemented independently. This, however, results in low spectrum utilization and expensive hardware overhead. Fortunately, integrated sensing and communications (ISAC), a key technique for next-generation wireless networks, is capable of simultaneously performing wireless communications and radar sensing based on the common spectrum resource and hardware infrastructures \cite{liu2022integrated}. With ISAC, the ground base station (GBS) provides communication and navigation services for authorized aircraft, while sensing the low-altitude airspace to monitor unauthorized targets.

\subsection{Related Work}
Currently, there have been substantial bodies of works studying the integration of aircraft (specifically UAVs) and ISAC systems, which can be divided into two main paradigms according to the role that UAVs play \cite{zhang2021uav, zhang2020age, wang2020constrained, chang2022integrated, ding2023multi, bayessa2024joint, yu2024security, wu2023uavs, hu2022trajectory, liu2023sensing, liu2024uav, liu2024secure, zhang2024secure, jing2024isac, meng2022throughput, meng2022uav, lyu2022joint, wu2024joint, deng2023beamforming, van2024joint, wu2023interplay, zhang2024joint, zhang2022trajectory, xie2024distributed, chen2024drl, cho2024enhancing, fontanesi2024deep, qin2023deep, dai2024joint, cui2023specific, cheng2024networked}.
The first paradigm is referred to as UAV-assisted ISAC services, where each UAV operates as a new aerial platform (e.g., aerial base station or relay) to serve ground users. For example, towards the goal of minimizing the age of information of sensed data, a joint extremum principles and dynamic programming scheme was proposed in \cite{zhang2021uav} to optimize the UAV power and trajectory, whereas in \cite{zhang2020age}, a successive convex approximation (SCA) iterative method was used to refine the UAV trajectory and time allocation for both communication and sensing. These two works, however, were merely concerned about the communication performance, while ignoring the sensing requirement. By contrast, the works \cite{wang2020constrained, chang2022integrated, ding2023multi, bayessa2024joint, yu2024security} considered both simultaneously. Specifically, in \cite{wang2020constrained}, Wang \textit{et al.} investigated a dual-function multi-UAV wireless network, in which the sum-rate was maximized by jointly optimizing the UAV location, transmit power, and user association under the Cram\'{e}r-Rao bound constraint. In \cite{chang2022integrated}, Chang \textit{et al.} proposed a joint scheduling strategy for control, communication, and sensing in millimeter-wave UAV networks, such that the communication and sensing requirements can be met while maintaining good control performance. By jointly employing the alternating optimization and SCA method, the authors of \cite{ding2023multi} proposed a joint UAV deployment and power control scheme to maximize the minimum detection probability, whilst the authors of \cite{bayessa2024joint} optimized the deployment and joint communication and sensing precoder to balance communication rate and sensing accuracy. Besides, the potential of UAVs for physical layer security provisioning in ISAC systems was explored in \cite{yu2024security}.

The aforementioned research, however, mainly focused on the ISAC in quasi-static UAV scenarios, without exploiting the controllable  mobility over the three-dimensional (3D) space. To explore the advantages of UAV mobility, literature \cite{wu2023uavs, hu2022trajectory, liu2023sensing, liu2024uav, liu2024secure, zhang2024secure, jing2024isac, meng2022throughput, meng2022uav, lyu2022joint, wu2024joint, deng2023beamforming, van2024joint} studied the UAV trajectory optimization for ISAC applications. To be specific, by optimizing the UAV trajectory, the real-time secrecy rate was maximized in \cite{wu2023uavs}, while the UAV propulsion consumption was minimized in \cite{hu2022trajectory}. Furthermore, in \cite{liu2023sensing, liu2024uav, liu2024secure, zhang2024secure, jing2024isac, meng2022throughput, meng2022uav, lyu2022joint, wu2024joint, deng2023beamforming, van2024joint}, other variables are jointly optimized with the UAV trajectory, e.g., UAV transmit power and user scheduling \cite{liu2023sensing, liu2024uav, liu2024secure, zhang2024secure}, bandwidth allocation for users \cite{jing2024isac}, UAV-user association \cite{meng2022throughput}, UAV transmit beamforming \cite{meng2022uav, lyu2022joint, wu2024joint, deng2023beamforming}, and offloading task size \cite{van2024joint}. These above works were dedicated to the case with a single UAV and thus exempted from the UAV collision-avoidance constraint. On the contrary, investigations \cite{wu2023interplay} and \cite{zhang2024joint} focused on a multi-UAV-assisted ISAC network, where the trajectory of individual UAVs and their association with users are jointly optimized subject to the minimum collision avoidance distance. In addition, to circumvent the high computational complexity and poor adaptability of conventional optimization methods, several intelligent UAV-aided ISAC schemes were developed in \cite{zhang2022trajectory, xie2024distributed, chen2024drl, cho2024enhancing, fontanesi2024deep, qin2023deep, dai2024joint}.

The inherent limitation of the first paradigm is that it is not applicable to terrestrial ISAC oriented towards LAE, where authorized UAVs (as communication users) with missions (e.g., cargo deliver) and unauthorized targets (as sensing targets) are located at low altitudes in the 3D space. Instead, the second paradigm, which focuses on the use of GBS to provide aircraft with ISAC services, aligns better with LAE-oriented ISAC. Currently, there are few studies on this paradigm \cite{cui2023specific}, \cite{cheng2024networked}. In particular, the authors of \cite{cui2023specific} considered a system with one GBS and multiple UAVs, where a novel dual identity association-based ISAC approach was designed to enable fast and accurate beamforming towards different UAVs. Unlike \cite{cui2023specific}, the authors of \cite{cheng2024networked} studied a networked ISAC system, where multiple GBSs cooperatively transmit unified ISAC signals to communicate with multiple authorized UAVs and concurrently detect unauthorized targets. In prior works, the unauthorized target is presumed to be static, and the iterative alternating algorithm that focuses on one instantaneous optimization problem is adopted to optimize different variables in a separate manner. In practical LAE-oriented ISAC systems, however, (i) the locations of unauthorized targets are highly dynamic; (ii) the channel state information (CSI) is time-correlated since the next locations of the target and UAVs depend on their current locations; and (iii) the beamforming and trajectory are jointly optimized subject to UAVs' flight missions that last for a long period. As a result, \textit{the LAE-oriented ISAC problem is a typical long-term optimization problem}, in which all variables should be jointly optimized with respect to the average system performance during the flight mission.

Such a problem can be transformed and solved under the Markov decision process (MDP) model, in which an agent continuously optimizes its decisions towards the given long-term objective. Generally, an efficient solution for MDP can be derived through a simple dynamic programming algorithm, if the accurate state transition (e.g., the CSI transition) of the system is available \cite{sutton2018reinforcement}. However, in practical LAE-oriented ISAC systems, the mobility model of the target is difficult to obtain, in the sense that the state transition of the system is unknown. Fortunately, model-free deep reinforcement learning (DRL) excels at finding an efficient strategy for MDP in unknown environments, through a large number of trial-and-error interactions and reward-decision iterations \cite{mnih2015human}. In this regard, \textit{DRL is a promising candidate for solving the long-term optimization problem in LAE oriented-ISAC systems}. So far, DRL has shown its powerful capability in solving a wide range of sequential decision-making problems, e.g., robot control, game playing, and wireless communications \cite{ye2021multi}.
To the best of our knowledge, DRL has not been explored yet in existing works to design the LAE-oriented ISAC scheme.


\subsection{Contributions}
In this paper, we investigate an LAE-oriented ISAC system with the objective of maximizing the \textit{expected} communication sum-rate of all UAVs over a given flight period, by jointly optimizing the GBS's beamforming and authorized UAVs' trajectories. In addition, the constraints, including the average sensing signal-to-noise ratio (SNR) requirement for monitoring the mobile target, the UAVs' flight missions, the UAV collision avoidance, and the GBS's maximum transmit power, are considered. Overall, the main contributions of this paper are summarized below:



1) A problem is formulated as a specific MDP model termed episode task \cite{sutton2018reinforcement} to jointly optimize the GBS's beamforming and UAVs' trajectories, where each flight period is considered as an episode and the flight mission corresponds to the episode task. To find an efficient solution, we propose an ISAC scheme capable of supporting LAE, referred to as Deep LAE-ISAC (DeepLSC), by exploiting the DRL technique. The underpinning algorithm in DeepLSC is deep deterministic policy gradient (DDPG) \cite{lillicrap2015continuous}, since the formulated problem involves continuous control variables. The salient advantages of DeepLSC are that it (i) does not require the prior mobility information of the target and (ii) well caters the long-term optimization objective of the LAE-oriented ISAC system.

2) A new action selection policy, termed constrained noise-exploration policy, is developed for DeepLSC to guarantee the flight mission and maximum power constraints. The gist is to (i) introduce a scaling factor to refine the beamforming from the conventional noise-exploration policy \cite{lillicrap2015continuous} and (ii) decide whether to follow the trajectory decision of the conventional noise-exploration policy or choose straight flight based on the real-time locations of UAVs. In addition, we judiciously design a reward function for action evaluation, which incorporates the communication sum-rate, average sensing SNR requirement, and collision avoidance into the learning process.

3) A new hierarchical experience replay mechanism is proposed for DeepLSC to train the deep neural network (DNN). Specifically, in the traditional experience replay mechanism \cite{lin1992self}, since different experiences generated within a specific episode are collected and utilized in a separate manner, they cannot be guaranteed to appear together during training. DeepLSC, however, is tailored for episode tasks. To efficiently learn from the episode task, the hierarchical experience replay mechanism employs the experience sets, each containing all experiences within an episode, to jointly train the DNN.

4) We devise a symmetric experience augmentation mechanism to promote the convergence speed of DeepLSC. To be specific, DeepLSC is an online learning scheme, in the sense that the agent (i.e., the GBS) needs a lot of trial and error with the environment to generate sufficient experiences for training. Consequently, an efficient joint beamforming and trajectory policy takes a long time to learn. To circumvent this issue, the symmetric experience replay mechanism generates more new experience sets by simultaneously permuting the indexes of all variables, based on a prior experience set.

5) Extensive experimental results show that DeepLSC outperforms various schemes including the DeepLSC-CNE (i.e., DeepLSC with the conventional noise-exploration policy \cite{lillicrap2015continuous}), DeepLSC-CER (i.e., DeepLSC with the conventional experience replay mechanism \cite{lin1992self}), and AC2 (i.e., the actor-critic algorithm \cite{sutton2018reinforcement} with the constrained noise-exploration policy), in terms of both communication and sensing performance. Compared with DeepLSC-w (i.e., DeepLSC without symmetric experience augmentation), DeepLSC converges faster. Furthermore, under different simulation setups, DeepLSC is more robust than these baselines.

\subsection{Organization and Notations}
\textit{Organization:} The remainder of this paper is below. Section \ref{1} describes the system model. Section \ref{ProblemForTransform} presents the problem formulation and transformation. The DeepLSC scheme is detailed in Section \ref{DeepLSCFramework}, and numerical results are discussed in Section \ref{4}. Finally, Section \ref{5} concludes this paper.

\textit{Notations:} Vectors (matrices) are denoted by boldface lower (upper) case letters, $\mathbb{C}^{N{\times}M}$ represents the space of $N{\times}M$ complex matrices, $|\cdot|$ represents the absolute value, $||\cdot||_2$ represents the 2-norm, $\mathbb{E}[\cdot]$ represents the statistical expectation, while $\text{Tr}(\cdot)$, $(\cdot)^T$, and $(\cdot)^H$ represent the trace, transpose, and conjugate transpose, respectively.

\section{System Model}\label{1}

\begin{figure}[t]
	\centering
	\includegraphics[scale=0.45]{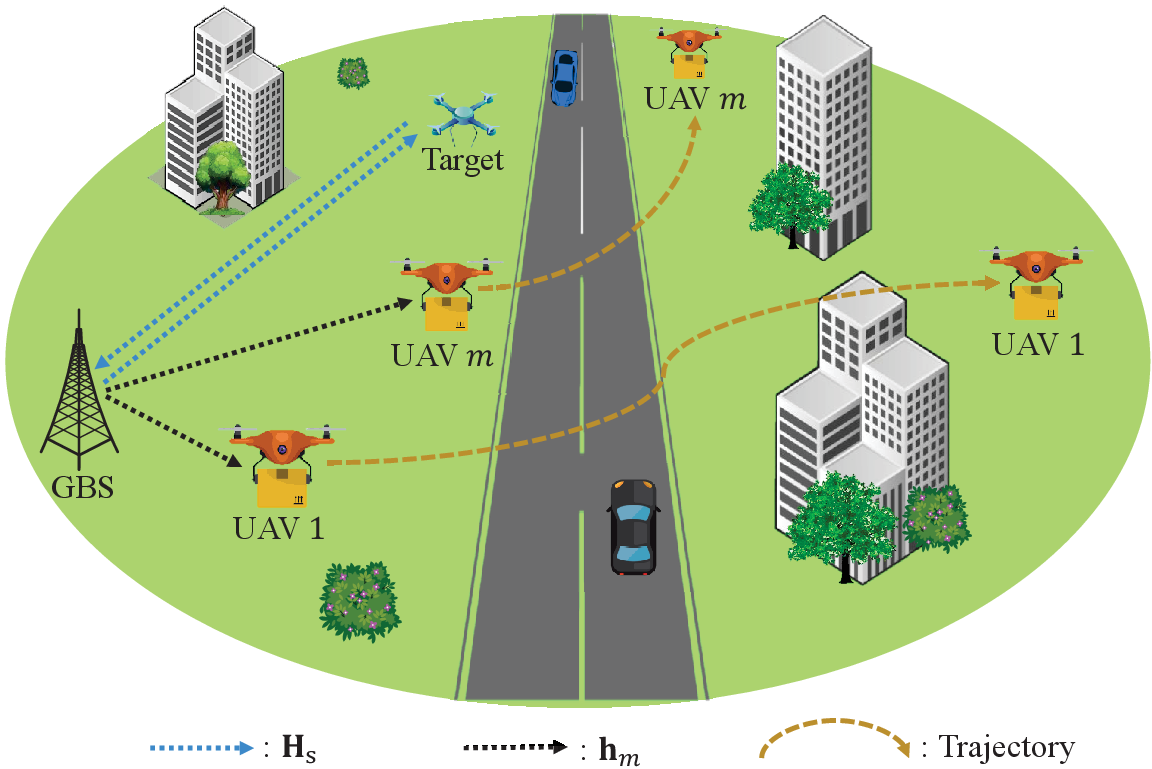}
	\caption{LAE-oriented ISAC systems. }
	\label{system1}
\end{figure}

Consider an LAE-oriented ISAC system shown in Fig. \ref{system1}, where one GBS equipped with $N$ transmit/receive-antennas serves $M$ single-antenna authorized UAVs for downlink communications and simultaneously detects a low-altitude target (e.g., unauthorized UAV). The location of the GBS is represented by the 3D Cartesian coordinate $(\mathbf{b}, 0)$ with $\mathbf{b}=[\hat{x},\hat{y}]$ denoting the horizontal coordinate, and the location of the target is $(\mathbf{g}, H_{\text{Tar}})$, wherein $\mathbf{g}=[\tilde{x},\tilde{y}]$ and $H_{\text{Tar}}>0$ denote the horizontal and vertical coordinates, respectively. The UAVs are indexed by $m\,{\in}\,\mathcal{M}=\{1,\,2,\,\cdots,\,M\}$, and each UAV $m$ flies at a pre-assigned altitude $H_m$ with the mission of transporting cargo from a source to a destination within a given time duration of ${\Delta}_{\text{T}}$. In other words, the initial and final locations of UAV $m$ are predetermined as $(\mathbf{u}_{m}^{\text{I}}, H_{m})$ and $(\mathbf{u}_{m}^{\text{F}}, H_{m})$, respectively, wherein $\mathbf{u}_{m}^{\text{I}}=[x_{m}^{\text{I}}, y_{m}^{\text{I}}]$ and $\mathbf{u}_{m}^{\text{F}}=[x_{m}^{\text{F}}, y_{m}^{\text{F}}]$.

\subsection{UAV/Target Trajectory Model}\label{UTM}
We discretize the duration $\Delta_{\text{T}}$ into $T$ time slots, indexed by $t\,{\in}\,{\mathcal{T}}=\{1,\,2,\,\cdots,\,T\}$. The slot duration $\Delta_{\text{t}}=\Delta_{\text{T}}/T$ is sufficiently small, such that the locations of the target and UAVs can be assumed as unchanged within each time slot.

\textit{1) UAV Flight Model:} Within a specific time period $\Delta_{\text{T}}$, the trajectory of UAV $m$ is a $(T+1)$-length sequence, i.e., $\{(\mathbf{u}_m^{\text{I}}, H_{m}), (\mathbf{u}_m(t), H_{m})|t=1,\,2,\,\cdots,\,T\}$ with $\mathbf{u}_m(t)\,{\triangleq}\,$ $[x_m(t),$ $y_m(t)]$. We assume that UAV $m$ flies at a constant speed $v_m$. Thus, its flight distance in one time slot is given as $v_m{\Delta}_{t}$, leading to the following constraint:
\begin{equation}
{\lVert}\mathbf{u}_m(t+1)-\mathbf{u}_m(t){\rVert}_2=v_m{\Delta}_{\text{t}},\,\forall\,m\,{\in}\,\mathcal{M},\,t\,{\in}\,\mathcal{T}.\label{UAVMove}
\end{equation}

In addition, the initial and final locations of $M$ UAVs (i.e., the flight mission constraints) are given by
\begin{equation}
\mathbf{u}_m(0)=\mathbf{u}_{m}^{\text{I}}\,\,\,\,\text{and}\,\,\,\,\mathbf{u}_m(T)=\mathbf{u}_{m}^{\text{F}},\,\forall\,m\,{\in}\,\mathcal{M}.\label{UAVInitialFinal}
\end{equation}

To avoid collision among different UAVs, the following collision avoidance constraints are enforced:
\begin{align}\nonumber
{\lVert}\mathbf{u}_m(t) - \mathbf{u}_i(t){\rVert}_2^2 &+ (H_m - H_i)^2 \geq D_{\text{min}}^2,\\
&\forall\,m,i\,{\in}\,\mathcal{M},\,m\,{\neq}\,i,\,t\,{\in}\,\mathcal{T},\label{UAVCollision}
\end{align}
where $D_{\text{min}}$ denotes the minimum allowed distance between any two UAVs. Similarly, each UAV also needs to maintain a minimum distance from the target to avoid collision, i.e.,
\begin{align}\nonumber
{\lVert}\mathbf{u}_m(t) - \mathbf{g}(t){\rVert}_2^2 + (H_m - &H_{\text{Tar}}(t))^2 \geq D_{\text{min}}^2,\\
&\forall\,m\,{\in}\,\mathcal{M},\,t\,{\in}\,\mathcal{T}.\label{UAVTarget}
\end{align}

\textit{2) Target Mobility Model:} For simplify, we assume that the target moves at a constant speed $v_{\text{Tar}}$ and its flight distance within one time slot is $v_{\text{Tar}}{\Delta}_{\text{t}}$. To better model the mobility of a realistic target, a Gauss-Markov process is introduced to capture the temporal correlation of the movement direction \cite{tabassum2019fundamentals}. Let $\phi(t)$ and $\varphi(t)$ denote the azimuth and elevation angles at which the target moves at time slot $t$, respectively. They are modeled as
\begin{equation}\label{MoveDirection}
  \begin{cases}
 \phi(t)=\mu_{\text{a}} {\phi(t-1)} + (1-\mu_{\text{a}}){\xi_{\text{a}}} + \sqrt{1-{\mu_{\text{a}}}^2} \hat{\phi}(t),\\
 \varphi(t)=\mu_{\text{e}} {\varphi(t-1)} + (1-\mu_{\text{e}}){\xi_{\text{e}}} + \sqrt{1-{\mu_{\text{e}}}^2} \hat{\varphi}(t),
  \end{cases}
\end{equation}
where $\mu_{\text{a}}\,{\in}\,[0,1]$ and $\mu_{\text{e}}\,{\in}\,[0,1]$ are the time correlation coefficients, modulating the degree of temporal dependency. For example, $\mu_{\text{a}}=0$ indicates that the target moves at a fully random azimuth angle, while $\mu_{\text{a}}=1$ indicates that the horizontal movement direction is unchanged. Parameters ${\xi_{\text{a}}}$ and ${\xi_{\text{e}}}$ are asymptotic means of $\phi(t)$ and $\varphi(t)$, respectively, as $t{\rightarrow}\infty$. Parameters $\hat{\phi}(t)$ and $\hat{\varphi}(t)$ are independent, uncorrelated, and stationary Gaussian processes with $\mathcal{N}(0,\sigma_{\phi}^2)$ and $\mathcal{N}(0,\sigma_{\varphi}^2)$, respectively, wherein $\sigma_{\phi}$ and $\sigma_{\varphi}$ correspond to the asymptotic standard deviations.

Hence, given $\phi(t)$ and $\varphi(t)$, the location of the target at time slot $t+1$ can be obtained as
    \begin{equation}
  \begin{cases}
 \mathbf{g}(t+1)=\mathbf{g}(t) + v_{\text{Tar}}{\Delta_{\text{t}}}{\cos}\left( \varphi(t) \right) \left[ {\cos}\left( \phi(t) \right), {\sin}\left( \phi(t) \right) \right],\\
 H_{\text{Tar}}(t+1)=H_{\text{Tar}}(t)+v_{\text{Tar}}{\Delta_{\text{t}}}{\sin}\left( \varphi(t) \right).\label{TargetMove}
  \end{cases}
\end{equation}

\subsection{Channel Model}\label{ChannelModel}
Denote $\mathbf{h}_{\text{s}}(t)\,{\in}\,{\mathbb{C}}^{N{\times}1}$ by the channel vector between the GBS and the target at time slot $t$. The bidirectional channel matrix that the signal experiences from forward transmission to echo can be represented as $\mathbf{H}_{\text{s}}(t)=\mathbf{h}_{\text{s}}(t) \mathbf{h}_{\text{s}}^T(t)\,{\in}\,{\mathbb{C}}^{N{\times}N}$. Furthermore, let $\mathbf{H}_{\text{c}}(t)=$ $[\mathbf{h}_1(t),\,\mathbf{h}_2(t),\,\cdots,\,\mathbf{h}_M(t)]\,{\in}\,{\mathbb{C}}^{N{\times}M}$ denote the channel matrix between the GBS and all UAVs at time slot $t$, wherein $\mathbf{h}_m(t)\,{\in}\,{\mathbb{C}}^{N{\times}1}$ is the channel vector between the GBS and UAV $m$.

Due to the relatively high altitude of the target and UAVs, there generally exists a strong line-of-light (LoS) link between the GBS and them. As such, the air-ground communication links are dominated by the LoS channel. For example, $\mathbf{h}_m(t),\,\forall\,m\,{\in}\,{\mathcal{M}}$ is represented as
\begin{equation}
\mathbf{h}_m(t) = \sqrt{\omega_m(t)} {\mathbf{c}}\left({\psi_m}(t)\right),\label{ChannelLoS}
\end{equation}
where
\begin{equation}
{\omega_m}(t) = {L_0} \frac{D_0}{{({\lVert}\mathbf{b}-\mathbf{u}_m(t){\rVert}_2^2+H_m^2})^{\varsigma_m}}\nonumber
\end{equation}
is the path loss with $L_0$ as the path loss constant at reference distance $D_0$, and $\varsigma_m$ as the path loss exponent. Furthermore, ${\mathbf{c}}\left( {\psi_m}(t) \right) = [ 1, e^{\jmath2{\pi}\frac{d}{\lambda}\cos{\psi_m(t)}},\cdots,e^{\jmath2{\pi}\frac{d}{\lambda}(N-1)\cos{\psi_m(t)}} ]^T$ is the steering vector with ${\psi_m}(t)$ being the angle of departure (AoD) at the GBS, given by
\begin{equation}
\psi_m(t) = \arccos \frac{H_m}{\sqrt{{\lVert}\mathbf{b}-\mathbf{u}_m(t){\rVert}_2^2+H_m^2}}.
\end{equation}
A similar channel model is adopted for $\mathbf{h}_{\text{s}}(t)$.

\subsection{Signal Model}\label{SignalModel}
In each time slot $t$, the transmit signal of the GBS is a weighted sum of communication symbols and radar probing signals \cite{liu2022integrated}, which is expressed as
\begin{equation}
\mathbf{x}(t) = \mathbf{W}_{\text{c}}(t) \mathbf{v}_{\text{c}}(t) + \mathbf{W}_{\text{s}}(t) \mathbf{v}_{\text{s}}(t),
\end{equation}
where $\mathbf{v}_{\text{c}}(t)\,{\in}\,{\mathbb{C}}^{M{\times}1}$ is the communication symbol with $\mathbb{E}\{\mathbf{v}_{\text{c}}(t)\mathbf{v}_{\text{c}}^H(t)\}=\mathbf{I}_M$, $\mathbf{v}_{\text{s}}(t)\,{\in}\,{\mathbb{C}}^{N{\times}1}$ is the radar probing signal with $\mathbb{E}\{\mathbf{v}_{\text{s}}(t) \mathbf{v}_{\text{s}}^H(t)\}=\mathbf{I}_N$, $\mathbf{W}_{\text{c}}(t)\,{\in}\,{\mathbb{C}}^{N{\times}M}$ is the communication precoding matrix, and $\mathbf{W}_{\text{s}}(t)\,{\in}\,{\mathbb{C}}^{N{\times}N}$ is the radar precoding matrix. In addition, $\mathbf{v}_{\text{c}}(t)$ and $\mathbf{v}_{\text{s}}(t)$ are statistically independent, leading to $\mathbb{E}(\mathbf{v}_{\text{c}}(t) \mathbf{v}_{\text{s}}^H(t))=\mathbf{0}$. Denote $\mathbf{W}(t){\triangleq}\left[ \mathbf{W}_{\text{c}}(t), \mathbf{W}_{\text{s}}(t) \right]$ and $\mathbf{v}(t){\triangleq}\left[ \mathbf{v}_{\text{c}}^T(t), \mathbf{v}_{\text{s}}^T(t) \right]^T$. Thus, we have $\mathbf{x}(t) = \mathbf{W}(t) \mathbf{v}(t)$, and the power constraint is given by
\begin{equation}
\text{Tr} \left( \mathbf{W}_{\text{c}}^H(t) \mathbf{W}_{\text{c}}(t) \right) + \text{Tr} \left( \mathbf{W}_{\text{s}}^H(t) \mathbf{W}_{\text{s}}(t) \right)\leq\,{P_{\text{max}}},\,\forall\,t\,{\in}\,\mathcal{T}.
\end{equation}

\textit{1) Communication Model:} With the transmit signal $\mathbf{x}(t)$, the received signal at UAV $m$ is expressed as $y_m(t) = \mathbf{h}_{m}^T(t) \mathbf{x}(t) + {n_m}(t)$, where $n_m(t)\,{\sim}\,\mathcal{C}\mathcal{N}(0, {\sigma}_m^2)$ is the additive white Gaussian noise (AWGN), and ${\sigma}_{\text{m}}^2$ is the noise power. As a result, the communication signal-to-interference-and-noise ratio (SINR) at UAV $m$ is given by
\begin{equation}
\text{SINR}_m(t) = \frac{\left| \mathbf{h}_{m}^T(t) \mathbf{w}_m(t)\right| ^2}{\sum\limits_{k=1,\,k{\neq}m}^{N+M}\left| \mathbf{h}_{m}^T(t) \mathbf{w}_k(t)\right| ^2+{\sigma}_m^2},\label{SINR1}
\end{equation}
where $\mathbf{w}_m$ is the $m$-th column of $\mathbf{W}$. Accordingly, the sum-rate of all UAVs can be calculated by
\begin{equation}
\text{R}_{\text{total}}(t) = \sum_{m=1}^M \log_2 \left( 1+\text{SINR}_m(t)\right).\label{Rate1}
\end{equation}

\textit{2) Sensing Model:} As shown in Fig. \ref{system1}, the radar probing signal is reflected by the target to GBS via the direct and reflected/refracted links. Thus, the echo signal from the target to GBS is expressed as $\mathbf{y}_\text{s}(t) = \mathbf{H}_{\text{s}}(t) \mathbf{x}(t) + \mathbf{n}_{\text{B}}(t)$, wherein $\mathbf{n}_{\text{B}}(t)\,{\sim}\,\mathcal{C}\mathcal{N}(\mathbf{0}_N,$ ${\sigma}_{\text{b}}^2{\mathbf{I}_N})$ is the AWGN, and ${\sigma}_{\text{b}}^2$ is the noise power. Furthermore, the sensing SNR of the target is given by
\begin{align}\nonumber
\text{SNR}_{\text{Tar}}(t) &= \frac{\mathbb{E} \left\{ \left| \mathbf{H}_{\text{s}}(t) \mathbf{W}(t) \mathbf{v}(t)\right| ^2 \right\}}{{\sigma}_{\text{b}}^2}\\
&= \frac{\text{Tr} \left( \mathbf{W}^H(t) \mathbf{H}_{\text{s}}^H(t) \mathbf{H}_{\text{s}}(t) \mathbf{W}(t) \right)}{{{\sigma}_{\text{b}}^2}}.\label{SNR1}
\end{align}

\section{Problem Formulation and Transformation}\label{ProblemForTransform}
This section formulates the joint beamforming and trajectory optimization problem as a constrained sequential decision problem. Then, it is transformed to an unconstrained problem with a judiciously designed reward function.

\subsection{Problem Formulation}\label{ProblemFormulation}
Our objective is to maximize the expected sum-rate of all UAVs over $T$ time slots while satisfying the sensing SNR requirement, flight mission, collision avoidance, and maximum transmit power constraints. The communication beamforming $\mathbf{W}_{\text{c}}(t)$, sensing beamforming $\mathbf{W}_{\text{s}}(t)$, and trajectory $\mathbf{u}_m(t)$ of each UAV are jointly optimized. Mathematically, this optimization problem is formulated as
\begin{subequations}\label{pf}
\begin{align}
\max\limits_{\mathbf{W}_{\text{c}}(t),\,\mathbf{W}_{\text{s}}(t),\,\mathbf{u}_m(t)}\,\,&\mathbb{E} \left[ \sum\limits_{t=1}^{T}\text{R}_{\text{total}}(t) \right] \label{pf0}\\
\text{s.t.}\,\,&\eqref{UAVMove}, \eqref{UAVInitialFinal}, \eqref{UAVCollision}, \eqref{UAVTarget}, \eqref{TargetMove},\label{pf1}\\
&\mathbb{E} \left[ \frac{1}{T}\sum\limits_{t=1}^{T}\text{SNR}_{\text{Tar}}(t) \right] \,{\geq}\,{\Gamma}_{\text{min}},\label{pf2}\\
&\text{Tr} \left( \mathbf{W}_{\text{c}}^H(t) \mathbf{W}_{\text{c}}(t) \right) + \text{Tr} \left( \mathbf{W}_{\text{s}}^H(t) \mathbf{W}_{\text{s}}(t) \right)\nonumber\\
&\leq\,{P_{\text{max}}},\,\forall\,t\,{\in}\,\mathcal{T},\label{pf3}
\end{align}
\end{subequations}
with the expectation in \eqref{pf0} taken over varying target mobility, which is unknown to the GBS. Besides, constraint \eqref{pf1} specifies the flight constraints of the target and UAVs; constraint \eqref{pf2} guarantees the average SNR requirement $\Gamma_{\text{min}}$ for sensing the target \cite{mu2023uav}; and constraint \eqref{pf3} limits the transmit power of the GBS to $P_{\text{max}}$.

Typically, problem \eqref{pf} is a sequential decision-making problem, which can be transformed and solved under the MDP model \cite{sutton2018reinforcement}.
This corresponds to one particular MDP form, referred to as \textit{episode task} \cite{sutton2018reinforcement}, where the task ends in a terminal status (i.e., the final location of the UAV) that separates the agent-environment interactions into episodes with each containing $T$ time slots. Different episodes are independent of each other, and a new episode will start from the starting status (i.e., the initial location of the UAV) after each episode ends. Besides, within each episode, the agent has to satisfy the preset constraints \eqref{pf1}--\eqref{pf3}. Thus, problem \eqref{pf} is actually a \textit{constrained episode task}. Generally, an efficient solution that can handle this problem is hard to derive \cite{sutton2018reinforcement}, which therefore motivates us to transform problem \eqref{pf} into an unconstrained MDP problem.

\subsection{Unconstrained MDP Transformation}\label{UMDPT}
We define the basic MDP elements, including agent, environment, action, state, transition probability, and reward function, as follows.

\textit{1) Agent and Environment $\mathcal{O}$:} The GBS is considered as the agent, while the target, UAVs, and wireless channels are treated as the environment. In each time slot, the GBS first decides an efficient joint beamforming and trajectory strategy. Afterward, it performs the communication beamforming to provide data transmission and navigation for all UAVs, while monitoring the mobile target via the sensing beamforming.

\textit{2) Action Space $\mathcal{A}$:} The action space $\mathcal{A}$ is composed of all possible actions that could be potentially executed by the agent. In the light of \eqref{pf}, the agent needs to decide three types of sub-actions, including the communication beamforming $\mathbf{W}_{\text{c}}(t)$, the sensing beamforming $\mathbf{W}_{\text{s}}(t)$, and all UAVs' movement directions $\mathbf{a}_{\text{u}}(t)$, in each time slot. Specifically, the sub-action $\mathbf{a}_{\text{u}}(t)$ is represented as $[a_{\text{u},1}(t),\,a_{\text{u},2}(t),$ $\cdots,\,a_{\text{u},M}(t)]$, where $a_{\text{u},m}(t)$ is the movement direction of UAV $m$ at time slot $t$. With $a_{\text{u},m}(t)$, the horizontal location of UAV $m$ at time slot $t+1$ becomes $\mathbf{u}_m(t+1)=$ $[x_m(t)+$ $v_m{\Delta}_{\text{t}}\cos(a_{\text{u},m}(t)),\,y_m(t)+v_m{\Delta}_{\text{t}}\sin(a_{\text{u},m}(t))]$. In $\mathbf{W}_{\text{c}}(t)$, $\mathbf{W}_{\text{s}}(t)$, and $\mathbf{a}_{\text{u}}(t)$, each element is a continuous variable to be optimized by the agent.
Overall, the action of the agent at time slot $t$ can be denoted by $\mathbf{A}(t) = \left( \mathbf{W}_{\text{c}}(t), \mathbf{W}_{\text{s}}(t), \mathbf{a}_{\text{u}}(t) \right)$.

\textit{3) State Space $\mathcal{S}$:} The state space $\mathcal{S}$ is a set of all possible states, each of which contains enough useful information for decision-making. To be specific, at time slot $t$, the observable state $\mathbf{S}(t)$ is determined by the communication CSI ${\mathbf{H}}_{\text{c}}(t)$, the sensing CSI ${\mathbf{H}}_{\text{s}}(t)$, and all UAVs' locations ${\mathbf{U}}(t)=[{\mathbf{u}}_1(t),{\mathbf{u}}_2(t),\cdots,{\mathbf{u}}_M(t)]$, which can be represented as
\begin{align}
\mathbf{S}(t) = {\big[} {\mathbf{H}}_{\text{c}}(t), {\mathbf{H}}_{\text{s}}(t), {\mathbf{U}}(t) {\big]}.\label{state1}
\end{align}


\textit{4) Transition Probabilities $\mathcal{P}$:} The transition probability, denoted by $p \left(\mathbf{S}(t+1)=\mathbf{S}'|\mathbf{S}(t)=\mathbf{S},\,\mathbf{A}(t)=\mathbf{A}\right)$, describes the probability that the state transits from $\mathbf{S}$ to $\mathbf{S}'$ after action $\mathbf{A}$ is conducted \cite{sutton2018reinforcement}. Since the mobility information of the target is not available, the transition probability $p \left(\mathbf{S}'|\mathbf{S},\,\mathbf{A}\right)$ is unknown. As a result, problem \eqref{pf} is a partially observable MDP problem.

\textit{5) Reward Function $\mathcal{R}$:} The reward function, denoted by $r(t+1)=\mathcal{R}(\mathbf{S}(t),\mathbf{A}(t))$, presents the award or penalty to evaluate how good $\mathbf{A}(t)$ is under $\mathbf{S}(t)$. In the design of the reward function, we address constraints \eqref{UAVCollision}, \eqref{UAVTarget}, and \eqref{pf2}, apart from the optimization objective \eqref{pf0}. In Section \ref{CNEP}, we will detail how to guarantee constraints \eqref{UAVInitialFinal} and \eqref{pf3} from the perspective of action selection.
\begin{itemize}
    \item First, considering only the optimization objective of maximizing the sum-rate of all UAVs, the reward is given by $ r(t+1)={\text{R}_{\text{total}}}(t)$.
    \item Second, the agent needs to be aware of its unwise decisions, i.e., the collision avoidance constraint \eqref{UAVCollision} or \eqref{UAVTarget} cannot be met. Thus, when a collision occurs at time slot $t$, the corresponding reward $r(t+1)$ should be a penalty $- \delta_1$, wherein $\delta_1 > 0$ is the penalty coefficient corresponding to the collision avoidance constraint, balancing the utility and cost \cite{ye2021multi}. The larger $\delta_1$ is, the more the agent attaches importance to the given constraint.
    \item Third, according to constraint \eqref{pf2}, the agent needs to satisfy the preset sensing SNR requirement ${\Gamma}_{\text{min}}$. In this regard, if the average sensing SNR over the whole flight period (i.e., episode)\footnote{This paper uses the terms ``flight period'' and ``episode'' interchangeably.} is less than ${\Gamma}_{\text{min}}$, a penalty $-{\delta_2}({\Gamma}_{\text{min}} - {\sum_{t=1}^T}{\text{SNR}_{\text{Tar}}}(t)/T)$ should be added to $r(t+1)$, wherein $\delta_2 > 0$ is the penalty coefficient corresponding to the sensing SNR requirement constraint. As such, the agent can be encouraged to optimize its strategy, so as to meet the average SNR requirement for sensing the target in subsequent episodes.
\end{itemize}

Overall, the reward function of the agent is designed as \eqref{rewardfunction1}. Note that since ${\sum_{t=1}^T}{\text{SNR}_{\text{Tar}}}(t)/T$ is not revealed until the end of each episode, the rewards $\{r(t+1)|t=1,\,2,\,\cdots,\,$ $T\}$ are only available at the agent in time slot $t=T$.

\begin{figure*}
\begin{equation}
r(t+1)=
  \begin{cases}
 - \delta_1 - {\delta_2} \left( {\Gamma}_{\text{min}} - \frac{1}{T}{\sum\limits_{t=1}^T}{\text{SNR}_{\text{Tar}}}(t) \right),\,&\text{if \eqref{UAVCollision} or \eqref{UAVTarget} is not met and \eqref{pf2} is not met},\\
 - \delta_1,\,&\text{if \eqref{UAVCollision} or \eqref{UAVTarget} is not met but \eqref{pf2} is met},\\
 {\text{R}_{\text{total}}}(t)-{\delta_2} \left( {\Gamma}_{\text{min}} - \frac{1}{T}{\sum\limits_{t=1}^T}{\text{SNR}_{\text{Tar}}}(t) \right),\,&\text{if \eqref{UAVCollision} and \eqref{UAVTarget} are met but \eqref{pf2} is not met},\\
 {\text{R}_{\text{total}}}(t),\,&\text{if \eqref{UAVCollision}, \eqref{UAVTarget}, and \eqref{pf2} are all met}.\label{rewardfunction1}
  \end{cases}
\end{equation}
\hrulefill
\end{figure*}

Thus far, we have reformulated problem \eqref{pf} as an unconstrained MDP problem. At each time slot, given the system state $\mathbf{S}(t)$, the agent makes an decision $\mathbf{A}(t)$ to interact with the environment by trial-and-error according to a policy $\pi(\mathbf{A}(t)|\mathbf{S}(t))$. Afterward, the environment feeds back a reward $r(t+1)$ to the agent, and enters a new state $\mathbf{S}(t+1)$ with probability $p \left(\mathbf{S}'|\mathbf{S},\,\mathbf{A}\right)$. Over the lifetime, the agent aims to search for an optimal policy ${\pi}^*$ (i.e., the optimal joint GBS's beamforming and UAVs' trajectories policy) to maximize the cumulative reward ${G(t)}=$ $\sum_{l = t}^{T} {r(l+1)}$ \cite{sutton2018reinforcement}. Furthermore, the action-value that reflects the expected cumulative reward at state $\mathbf{S}(t)$ given $\mathbf{A}(t)$ is chosen can be defined as
\begin{align}
{Q_\pi } \left({\mathbf{S}(t),\mathbf{A}(t)}\right)={\mathbb{E}_{\pi} }\left[ {{G(t)}|{\mathbf{S}(t)},{\mathbf{A}(t)}} \right].\label{QFunction}
\end{align}

In the above MDP problem, if the channel model is accurately known, the state transition probability $p \left(\mathbf{S}'|\mathbf{S},\,\mathbf{A}\right)$ is easy to obtain. At this time, this MDP problem can be well tackled by dynamic programming. However, the mobility model of the target is unavailable at the GBS, which motivates the use of model-free DRL techniques.


\section{DeepLSC Scheme}\label{DeepLSCFramework}
In this section, we present DeepLSC, a DRL-enabled ISAC scheme, to solve the formulated MDP problem. The underpinning algorithm in DeepLSC is DDPG \cite{lillicrap2015continuous}, and the major reasons why we favor DDPG are as follows: (i) Problem \eqref{pf} involves continuous control variables, whereas value-based DRL methods, e.g., deep Q-network (DQN) \cite{mnih2015human}, are limited to discrete control problems. (ii) Conventional policy-based approaches, e.g., policy gradient \cite{sutton1999policy}, are capable of solving continuous control issues, but they suffer from unstable learning and slow convergence. DDPG, which integrates the advantages of value-based and policy-based methods, can effectively circumvent the above challenges. Fig. \ref{DeepLSCFramework1} shows the DeepLSC framework, and the details are elaborated below.

\subsection{Actor-Critic Architecture}
\begin{figure*}[t]
	\centering
	\includegraphics[scale=0.5]{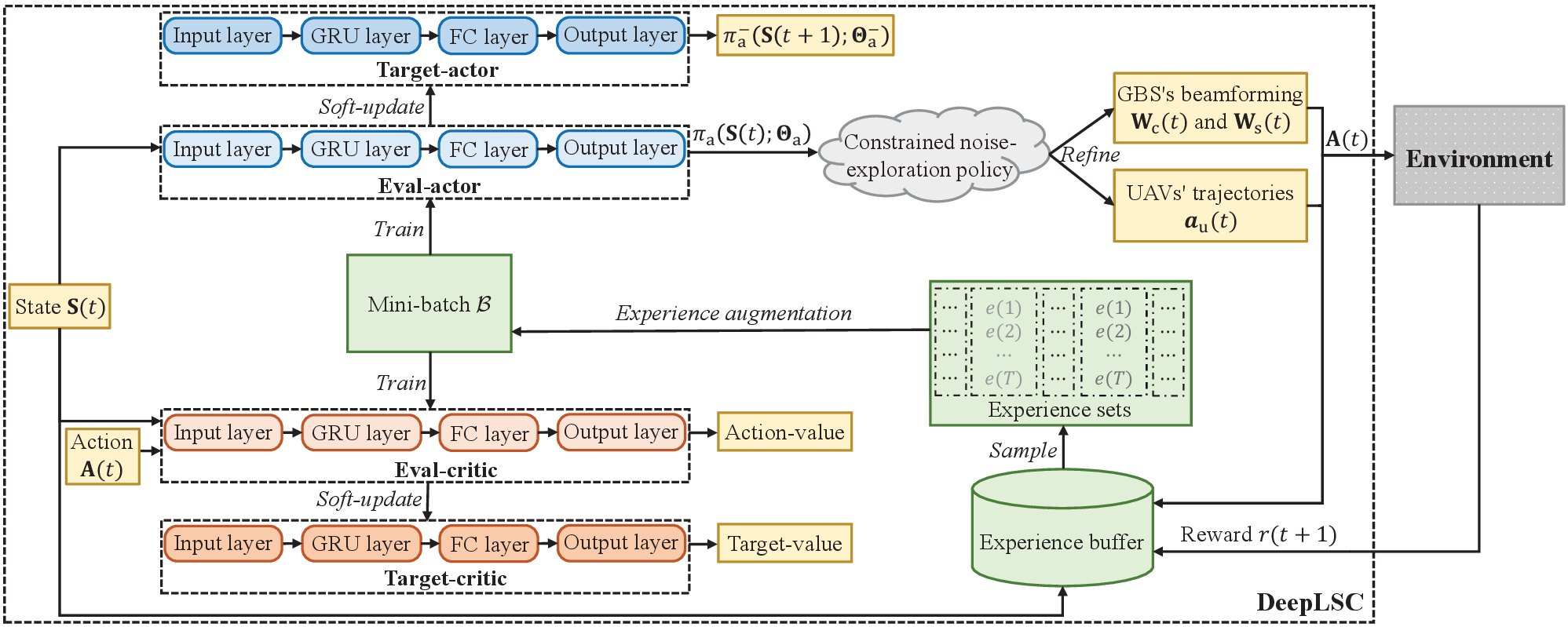}
	\caption{DeepLSC framework, including the execution phase and the training phase. During the execution phase, based on $\mathbf{S}(t)$, the eval-actor outputs the temporary joint beamforming and trajectory decision ${\pi}_{\text{a}}(\mathbf{S}(t),\mathbf{\Theta}_{\text{a}})$, which is refined by the constrained noise-exploration policy. During the training phase, some experience sets sampled from the experience buffer are augmented to form the mini-batch for training the actor-critic architecture. }
	\label{DeepLSCFramework1}
\end{figure*}

There are two types of DNNs in DeepLSC, referred to as actor and critic. Each of them consists of two DNNs with the same structure but different parameters, i.e., the eval-actor with parameter $\mathbf{\Theta}_{\text{a}}$, the target-actor with parameter $\mathbf{\Theta}_{\text{a}}^-$, the eval-critic with parameter $\mathbf{\Theta}_{\text{c}}$, and the target-critic with parameter $\mathbf{\Theta}_{\text{c}}^-$.
As shown in Fig. \ref{DeepLSCFramework1}, all DNNs share the same structure, i.e., one input layer, one gated recurrent unit (GRU) layer, one fully connected (FC) layer, and one output layer, but their roles are different.

\textit{1) Eval-Actor:} At each time slot, the eval-actor makes the \textit{temporary} joint beamforming and trajectory decision ${\pi}_{\text{a}} \left( \mathbf{S}(t); \mathbf{\Theta}_{\text{a}} \right)$ based on the state $\mathbf{S}(t)$. Specifically, since the DNN cannot process complex numbers, we divide ${\mathbf{H}}_{\text{c}}(t)$ and ${\mathbf{H}}_{\text{s}}(t)$ into the corresponding real and imaginary parts, and concatenate them together with $\mathbf{U}(t)$. After obtaining the preprocessed state $\mathbf{S}(t)$, the input layer delivers it into the GRU layer. Afterward, the GRU layer extracts the underlying temporal correlation from the input sequence and imports it into the FC layer for analysis.\footnote{Our previous work \cite{ye2021multi} has demonstrated the efficiency of GRU in addressing the problem with complex temporal correlations in wireless networks.} Since DNNs cannot generate complex numbers, it is not feasible to directly get a matrix with size $N{\times}M$ as $\mathbf{W}_{\text{c}}(t)$ and a matrix with size $N{\times}N$ as $\mathbf{W}_{\text{s}}(t)$. Thus, after receiving the processed information from the FC layer, the output layer generates ${\pi}_{\text{a}} \left( \mathbf{S}(t); \mathbf{\Theta}_{\text{a}} \right)$ containing three vectors, wherein the first vector with size $2NM$ is related to the temporary communication beamforming, the second vector with size $2N^2$ is related to the temporary sensing beamforming, and the last vector with size $M$ is the temporary trajectory decision. In particular, with the first vector in ${\pi}_{\text{a}} \left( \mathbf{S}(t); \mathbf{\Theta}_{\text{a}} \right)$, the temporary communication beamforming can be generated via element-wise concatenation, i.e., the first $NM$ elements are treated as the real parts of individual elements in the temporary communication beamforming and the remaining elements are the corresponding imaginary parts. The temporary sensing beamforming can also be generated in a similar way. On this basis, we design a new action selection policy to determine the \textit{real} joint beamforming and trajectory decision $\mathbf{A}(t)$, which will be detailed in Section \ref{CNEP}.

\textit{2) Eval-Critic:} The eval-critic is responsible for evaluating the action-value $Q(\mathbf{S}(t), \mathbf{A}(t); \mathbf{\Theta}_{\text{c}})$ of the pair $(\mathbf{S}(t), \mathbf{A}(t))$. In particular, after acquiring the action $\mathbf{A}(t)$ from the eval-actor, the agent feeds it together with $\mathbf{S}(t)$ to the eval-critic for action-value evaluation. To cater to the input of the DNN, the eval-critic preprocesses each element in both $\mathbf{S}(t)$ and $\mathbf{A}(t)$ in a similar way as the eval-actor. Then, the evaluated action-value will be used to compute the action gradient, i.e., ${{\nabla}_{\mathbf{A}(t)}} Q {\big(} {\mathbf{S}(t)}, \mathbf{A}(t); \mathbf{\Theta}_{\text{c}} {\big)}$, so as to train the eval-actor as detailed in Section \ref{ExperienceReplay}.

\textit{3) Target-Actor and Target-Critic:} The role of both the target-actor and the target-critic is to compute the target value of $Q(\mathbf{S}(t), \mathbf{A}(t); \mathbf{\Theta}_{\text{c}})$ when the eval-critic is trained. To further enhance the stability of the algorithm, the parameters of the eval-actor and the eval-critic are fully updated, whereas the parameters of the target-actor and the target-critic are partially updated via the soft-update approach in Section \ref{SoftUpdate}.

\subsection{Constrained Noise-Exploration Policy}\label{CNEP}
Given the optimization objective \eqref{pf}, the eval-actor aims to find the optimal policy of joint beamforming and trajectory ${\pi}_{\text{a}}^*$. Recall that in Section \ref{UMDPT}, constraints \eqref{UAVCollision}, \eqref{UAVTarget}, and \eqref{pf2} have been considered from the perspective of reward design. We now illustrate how to guarantee the maximum power constraint \eqref{pf3} and the flight mission constraint \eqref{UAVInitialFinal} from the perspective of action selection. Specifically, 
according to constraints \eqref{pf3} and \eqref{UAVInitialFinal}, the signal power used for communication and sensing cannot exceed the preset maximum value $P_{\text{max}}$, and the trajectory should ensure all UAVs to complete their flight missions within each episode. Given the temporary joint beamforming and trajectory decision ${\pi}_{\text{a}} \left( \mathbf{S}(t); \mathbf{\Theta}_{\text{a}} \right)$ from the eval-actor based on $\mathbf{S}(t)$, the conventional noise-exploration policy in DDPG \cite{lillicrap2015continuous} generates a refined decision by adding randomness, i.e.,
\begin{equation}
(\mathbf{a}_{\text{c}}(t),\,\mathbf{a}_{\text{s}}(t),\,\mathbf{a}_{\text{u}}(t))={\pi}_{\text{a}} \left( \mathbf{S}(t); \mathbf{\Theta}_{\text{a}} \right) + \left( \mathbf{d}_{\text{c}}(t),\,\mathbf{d}_{\text{s}}(t),\,\mathbf{d}_{\text{u}}(t) \right),\label{actionselection1}
\end{equation}
where $\mathbf{a}_{\text{c}}(t)$ and $\mathbf{a}_{\text{s}}(t)$ are vectors with size $2NM$ and $2N^2$, respectively. Vector $\mathbf{d}_{i}(t),\,\forall\,i{\in}\{\text{c},\,\text{s},\,\text{u}\}$, has the same size as $\mathbf{a}_{i}(t)$, each element of which is Gaussian distributed with zero mean and variance $\sigma_{i}^2(t)$. By doing so, the agent is encouraged to explore random actions for better solutions. Generally, too much exploration is unnecessary once an efficient solution is learned. Hence, we introduce \textit{a decay factor} $\kappa\,{\in}\,(0,1)$ to reduce $\sigma_{i}^2(t)$ over time. Denote the initial variances by $\sigma_{i,\text{init}}^2$, then $\sigma_{i}^2(t)$ at time slot $t$ can be represented as $\sigma_{i,\text{init}}^2{\kappa}^t$.


The conventional noise-exploration policy, however, only improves the exploration capability of the agent, while ignoring constraints \eqref{pf3} and \eqref{UAVInitialFinal}. To cope with this issue, we develop a new action selection policy for DeepLSC, referred to as \textit{constrained noise-exploration policy}. The gist is to (i) introduce a scaling factor to refine the beamforming decided by the conventional noise-exploration policy, so as to meet constraint \eqref{pf3}, and (ii) decide whether to use the conventional noise-exploration policy based on the real-time location of each UAV, so as to satisfy constraint \eqref{UAVInitialFinal}. The details are elaborated below.

\textit{1) Decision-Making of Communication and Sensing Beamforming:} As a first step, with $\mathbf{a}_{\text{c}}(t)$ and $\mathbf{a}_{\text{s}}(t)$, two matrices $\mathbf{A}_{\text{c}}(t)\,{\in}\,\mathbb{C}^{N{\times}M}$ and $\mathbf{A}_{\text{s}}(t)\,{\in}\,\mathbb{C}^{N{\times}N}$ can be generated via element-wise concatenation, i.e., the first $NM$ ($N^2$) elements of $\mathbf{a}_{\text{c}}(t)$ ($\mathbf{a}_{\text{s}}(t)$) are treated as the real parts of individual elements in $\mathbf{A}_{\text{c}}(t)$ ($\mathbf{A}_{\text{s}}(t)$) and the remaining elements are the corresponding imaginary parts. Thereafter, to obtain the communication beamforming $\mathbf{W}_{\text{c}}(t)$ and the sensing beamforming $\mathbf{W}_{\text{s}}(t)$ that satisfy the preset power constraint \eqref{pf3}, a scaling factor, defined as $\epsilon=$ ${P_{\text{max}}}/{\big(}\text{Tr} \left( \mathbf{A}_{\text{c}}^H(t) \mathbf{A}_{\text{c}}(t) \right) + \text{Tr} \left( \mathbf{A}_{\text{s}}^H(t) \mathbf{A}_{\text{s}}(t) \right)$, is introduced to refine $\mathbf{A}_{\text{c}}(t)$ and ${\mathbf{A}}_{\text{s}}(t)$ as follows:
\begin{equation}
 \mathbf{W}_{\text{c}}(t)=\sqrt{\epsilon}\mathbf{A}_{\text{c}}(t),\,\,\mathbf{W}_{\text{s}}(t)=\sqrt{\epsilon}{\mathbf{A}}_{\text{s}}(t).\label{actionselection2}
\end{equation}

\textit{2) Decision-Making of UAVs' Trajectories:} Within each episode, all UAVs' trajectories $\{\mathbf{a}_{\text{u}}(t)=[a_{\text{u},1}(t),\,a_{\text{u},2}(t),\,\cdots,$ $\,a_{\text{u},M}(t)]|t\,{\in}\,\mathcal{T}\}$ decided by the agent should fulfill the preset flight mission. For this purpose, the constrained noise-exploration policy that decides $a_{\text{u},m}(t)$ limits the movement directions of UAV $m$ in the last $T-t-1$ time slots within each episode:
at time slot $t$, if the minimum required number of time slots from the current location to reach the desired location is not less than $T-t-1$\footnote{Intuitively, to complete the flight mission, UAV $m$ only needs to fly in a straight line towards the desired location $\mathbf{u}_m^{\text{F}}$ when the number of its available time slots is equal to the minimum required number of time slots. However, there are some special cases. For example, the movement distance of UAV $m$ in each time slot is 1 m, the number of time slots available to UAV $m$ at time slot $t$ is 11, and at least 10 time slots are required to reach the desired location $\mathbf{u}_{\text{u},m}^{\text{F}}$. If UAV $m$ moves in the opposite direction of the desired location at time slot $t$, the number of time slots available to UAV $m$ at time slot $t+1$ will be 10 but at this time at least 11 time slots will be required to reach the desired location. Consequently, UAV $m$ will never complete the flight mission in the subsequent time slots. Therefore, in \eqref{actionselection3}, we relax $T-t$ to $T-t-1$.}, UAV $m$ has to fly in a straight line toward the desired location $\mathbf{u}_m^{\text{F}}$, i.e., $a_{\text{u},m}(t)=$ ``straight flight''. Within other time slots, the movement decision $a_{\text{u},m}(t)$ can be directly determined through the conventional noise-exploration policy, in the sense that the agent is not subject to constraint \eqref{UAVInitialFinal} and thus can derive more efficient movement directions for UAVs towards other optimization objectives. Overall, the movement decision of UAV $m$ at time slot $t$ is given as follows:
\begin{equation}
a_{\text{u},m}(t)=
  \begin{cases}
 a_{\text{u},m}(t),\,&\text{if $ \lceil \frac{\lVert \mathbf{u}_m(t)-\mathbf{u}_m^{\text{F}} \rVert_2}{{v_m}{\Delta_{\text{t}}}} \rceil < T-t-1$},\\
\text{straight\,\,flight},\,&\text{otherwise},\label{actionselection3}
  \end{cases}
\end{equation}
where $\lceil x \rceil$ represents the smallest integer not less than $x$.

\subsection{Hierarchical Experience Replay}\label{ExperienceReplay}
Since DeepLSC is developed for episode tasks, all experiences from different time slots within each episode should be jointly used to train both the eval-actor and the eval-critic. However, in the conventional experience replay mechanism of DDPG \cite{lin1992self}, the experiences generated in different time slots are separately gathered and utilized. In other words, during training, the experiences from different time slots within a specific episode cannot be guaranteed to appear in full. Consequently, the agent fails to learn an efficient joint beamforming and trajectory strategy in LAE-oriented ISAC systems. To fill this gap, a new \textit{hierarchical experience replay mechanism} is designed for DeepLSC, which makes use of experience sets each containing all experiences generated within each episode to train the eval-actor and the eval-critic.


As a first step, to store and reuse historical experiences, a first-input-first-output (FIFO) experience buffer $\mathcal{D}$ with size $D$ is used. As mentioned in Section \ref{UMDPT}, within each episode, the rewards $\{r(t+1)|t=1,2,\cdots,T\}$ are not available until time slot $T$. Thus, in time slot $T$ of each episode, after calculating the reward $r(t+1)$ via \eqref{rewardfunction1}, the agent collects the experience $e(t)$ in the form of $(\mathbf{S}(t), \mathbf{A}(t), r(t+1), \mathbf{S}(t+1))$, where $\mathbf{A}(t)=[\mathbf{W}_{\text{c}}(t),\mathbf{W}_{\text{s}}(t),\mathbf{a}_{\text{u}}(t)]$. With all experiences generated over the whole episode, a experience set $\mathcal{E}=\{e(t)|t=1,\,2,\,\cdots,\,T\}$ is formed and stored in $\mathcal{D}$.
During training, $N_{\text{e}}$ experience sets are randomly sampled from $\mathcal{D}$ to form the mini-batch $\mathcal{B}$. Afterward, the loss function for training the eval-critic is given as
\begin{align}
L(\mathbf{\Theta}_{\text{c}} ) = \frac{1}{N_{\text{e}}}\sum\limits_{\mathcal{E} \subset \mathcal{B}} {\Bigg(} \sum\limits_{t=1}^{T} \big( z(t+1) - Q \left( {\mathbf{S}(t)}, \mathbf{A}(t);\mathbf{\Theta}_{\text{c}} \right) \big) ^2 {\Bigg)},\label{loss1}
\end{align}
where $z(t+1)={r(t+1)} + Q \big( {\mathbf{S}(t+1)},{\pi}_{\text{a}}^- \big( \mathbf{S}(t+1);$ $\mathbf{\Theta}_{\text{a}}^- \big) ;\mathbf{\Theta}_{\text{c}}^- \big)$ is the target value related to the eval-actor, and ${\pi}_{\text{a}}^-$ is the policy of the target-actor. Furthermore, by performing the stochastic gradient descent (SGD) algorithm \cite{sutton2018reinforcement}, the parameter $\mathbf{\Theta}_{\text{c}}$ is updated as follows:
\begin{equation}
  \mathbf{\Theta}_{\text{c}}{\leftarrow}\mathbf{\Theta}_{\text{c}}-{\alpha_{\text{c}}}{{\nabla}_{\mathbf{\Theta}_{\text{c}}}}L(\mathbf{\Theta}_{\text{c}} ),\label{gradientdescent1}
\end{equation}
where $\alpha_{\text{c}}$ is the learning rate of $\mathbf{\Theta}_{\text{c}}$.
On the other hand, the parameter $\mathbf{\Theta}_{\text{a}}$ of the eval-actor can be updated via the sampled policy gradient algorithm \cite{lillicrap2015continuous}, i.e.,
\begin{align}\nonumber
\mathbf{\Theta}_{\text{a}}{\leftarrow}\mathbf{\Theta}_{\text{a}}-{\alpha_{\text{a}}}\frac{1}{N_{\text{e}}}\sum\limits_{\mathcal{E} \subset \mathcal{B}} {\Bigg(} \sum\limits_{t=1}^{T} &{\Big(} {{\nabla}_{\mathbf{A}(t)}} Q {\big(} {\mathbf{S}(t)},\mathbf{A}(t);\mathbf{\Theta}_{\text{c}} {\big)}\\
&{\times}{{\nabla}_{\mathbf{\Theta}_{\text{a}}}} {\pi}_{\text{a}} {\big(} \mathbf{S}(t); \mathbf{\Theta}_{\text{a}} {\big)} {\Big)} {\Bigg)},\label{gradientdescent2}
\end{align}
with $\alpha_{\text{a}}$ being the learning rate of $\mathbf{\Theta}_{\text{a}}$.

\subsection{Symmetric Experience Augmentation}
In DeepLSC, to gain enough experiences for DNN training, the agent needs to perform a lot of trial and error with the environment. Consequently, an efficient joint beamforming and trajectory strategy usually takes a long time to derive. If the experience set $\{(\mathbf{S}(t),\mathbf{A}(t),r(t+1),\mathbf{S}(t+1))|t=1,\,2,\,\cdots,\,T\}$ can be generated based on existing experience sets, the experience buffer $\mathcal{D}$ will effectuate experience augmentation and greatly boost the convergence speed.
Actually, in problem \eqref{pf}, the index of each UAV is artificially specified. Thus, in terms of the system performance, if the indices of different UAVs are permuted, the new optimization problem is equivalent to the original problem \cite{seress2003permutation}. For example, when $M=2$ and $N=2$, the communication CSI is given as
\begin{equation}
    \mathbf{H}_{\text{c}}(t)=\left[
    \begin{array}{cccc}
        \text{h}_{1,1}(t) & \text{h}_{1,2}(t)\\
        \text{h}_{2,1}(t) & \text{h}_{2,2}(t)\\
    \end{array}
    \right].
\end{equation}
where $\text{h}_{n,m}$ is the channel component of antenna $n$ to UAV $m$. If we permute the indexes of UAV 1 and UAV 2, the communication CSI will be transformed into
\begin{equation}
    \mathbf{H}_{\text{c}}(t)=\left[
    \begin{array}{cccc}
        \text{h}_{1,2}(t) & \text{h}_{1,1}(t)\\
        \text{h}_{2,2}(t) & \text{h}_{2,1}(t)\\
    \end{array}
    \right].
\end{equation}
At this time, if the indexes of UAV 1 and UAV 2 in the communication beamforming $\mathbf{W}_{\text{c}}(t)$, the sensing beamforming $\mathbf{W}_{\text{s}}(t)$, the UAVs' movement decisions $\mathbf{a}_{\text{u}}(t)$, and the sensing CSI $\mathbf{H}_{\text{s}}(t)$ are synchronously permuted, the new optimization problem will be equivalent to the original problem.

Inspired by this, we propose the \textit{symmetric experience augmentation mechanism}, and the main idea is to simultaneously exchange the indexes of all variables to generate more new experience sets.
Specifically, denote $\Upsilon=\{\{\upsilon_1,\,\upsilon_2,\,\upsilon_3,\,\cdots,\,\upsilon_M\},$ $\{\upsilon_2,\,\upsilon_1,\,\upsilon_3,\,\cdots,\,\upsilon_M\},\,\cdots,$ $\{\upsilon_M,\,\upsilon_{M-1},\,\upsilon_{M-2},\,\cdots,\,\upsilon_1\}\}$ by the symmetric group of finite integer set $\mathcal{F}=\{\upsilon_1,\,\upsilon_2,\,\cdots,\,\upsilon_M\}$, where group elements are bijection of $\mathcal{F}$ to itself. Let $\Omega_{\upsilon_i}$ denote the permutation of $\upsilon_i$. Then the permutation of individual elements in $\mathcal{F}$ can be represented as
\begin{equation}
    \Omega=\left(
    \begin{array}{cccc}
        \upsilon_1 & \upsilon_2 & \cdots & \upsilon_M\\
        \Omega_{\upsilon_1} & \Omega_{\upsilon_2} & \cdots & \Omega_{\upsilon_M}\\
    \end{array}
    \right),\label{permutation}
\end{equation}
where the first row is the elements of $\mathcal{F}$, and the second row is the corresponding permutation under $\Omega$. In the above example, the permutation is given as $\Omega = \begin{pmatrix} 1 & 2 \\ 2 & 1 \end{pmatrix}$. Consider a system with $M$ UAVs. Under all possible permutations $\Omega$, there will be a total of $M!$ group elements in $\Upsilon$. Besides, to distinguish the variables before and after the permutation, we superscript the permuted communication beamforming, sensing beamforming, UAVs' movement decisions, communication CSI, and sensing CSI by $\Omega$ as $\mathbf{W}_{\text{c}}^{\Omega}(t)$, $\mathbf{W}_{\text{s}}^{\Omega}(t)$, $\mathbf{a}_{\text{u}}^{\Omega}(t)$, $\mathbf{H}_{\text{c}}^{\Omega}(t)$, and $\mathbf{H}_{\text{s}}^{\Omega}(t)$, respectively.

Therefore, the experience sets used for DeepLSC training can be significantly enriched through various permutations. In particular, given a prior experience set $\big\{e(i)=(\mathbf{S}(i), \mathbf{A}(i), r(i+1), \mathbf{S}(i+1))|i=1,\,2,\,\cdots,\,T\big\}$ that is acquired via the agent-and-environment interaction, $M!-1$ new experience sets, i.e., ${\Big\{}{\big\{}e^{\Omega}(i)|i=1,\,2,\,\cdots,\,T{\big\}}|\Omega\,{\in}\,\Upsilon{\Big\}}$, can be generated based on the symmetric group $\Upsilon$. In each experience set $e^{\Omega}(i)$, the elements are given as
\begin{equation}
  \begin{cases}
 \mathbf{S}^{\Omega}(i)=[\mathbf{H}_{\text{c}}^{\Omega}(i), \mathbf{H}_{\text{s}}^{\Omega}(i), \mathbf{U}^{\Omega}(i)]\\
\mathbf{A}^{\Omega}(i)=[\mathbf{W}_{\text{c}}^{\Omega}(i), \mathbf{W}_{\text{s}}^{\Omega}(i), \mathbf{a}_{\text{u}}^{\Omega}(i)],\\
r^{\Omega}(i+1)=r(i+1),\\
\mathbf{S}^{\Omega}(i+1)=[\mathbf{H}_{\text{c}}^{\Omega}(i+1), \mathbf{H}_{\text{s}}^{\Omega}(i+1), \mathbf{U}^{\Omega}(i+1)].\label{NewExperience}
  \end{cases}
\end{equation}

The symmetric mapping, despite increasing the number of available experience sets, may lead to strong correlation among experience sets \cite{sutton2018reinforcement}. The consequence is that DeepLSC will fall into overfitting and converge to a sub-optimal joint beamforming and trajectory strategy. Actually, at the beginning of training, it is difficult to get enough experience sets. In this case, it is beneficial to augment existing experience sets based on the symmetric group $\Upsilon$. However, as the number of episodes increases, the experience sets from agent-and-environment interactions become miscellaneous. As a result, it is no longer necessary to use the symmetric mapping to enrich the available experience sets. Hence, we introduce a \textit{dynamic augmentation factor} $\zeta\,{\in}\,(0, 1)$ to adjust the number of augmented experience sets as follows:
\begin{equation}
    \Lambda(\omega)=\lfloor (M!-1){\zeta}^{\omega} \rfloor,
\end{equation}
where $\lfloor x \rfloor$ represents the largest integer less than $x$, and $\omega$ is the number of training times. During each training, $N_{\text{e}}$ experience sets, i.e., ${\Big\{}{\big\{}e(i,j)|i=1,\,2,\,\cdots,\,T{\big\}}|j=1,\,2,\,\cdots,\,N_{\text{e}}{\Big\}}$, are sampled from $\mathcal{D}$ in random, and each of which is augmented based on $\Lambda(\omega)$ random group elements (i.e., permutations $\Omega$) of the symmetric group $\Upsilon$. Thereafter, all augmented experience sets are merged with $N_{\text{e}}$ original experience sets to form the mini-batch $\mathcal{B}$ for training the eval-actor and eval-critic.


\subsection{Soft-Update for Target-Actor and Target-Critic}\label{SoftUpdate}
The parameters $\mathbf{\Theta}_{\text{a}}^-$ and $\mathbf{\Theta}_{\text{c}}^-$ of the target-actor and the target-critic are updated based on the parameters $\mathbf{\Theta}_{\text{a}}$ and $\mathbf{\Theta}_{\text{c}}$ of the eval-actor and the eval-critic.
A simple approach is to update the parameters of the target-actor and the target-critic by $\mathbf{\Theta}_{\text{a}}^-{\leftarrow}\mathbf{\Theta}_{\text{a}}$ and $\mathbf{\Theta}_{\text{c}}^-{\leftarrow}\mathbf{\Theta}_{\text{c}}$. However, this approach is not suitable for the continuous control task, since it may lead to large prediction errors of the target-actor and the target-critic \cite{lillicrap2015continuous}. To circumvent this issue, the soft-update approach was proposed in \cite{sutton2018reinforcement}, which updates both $\mathbf{\Theta}_{\text{a}}^-$ and $\mathbf{\Theta}_{\text{c}}^-$ as follows:
\begin{equation}
  \mathbf{\Theta}_{\text{a}}^- \leftarrow {\chi_{\text{a}}}\mathbf{\Theta}_{\text{a}}+(1-\chi_{\text{a}})\mathbf{\Theta}_{\text{a}}^-,\,
 \mathbf{\Theta}_{\text{c}}^- \leftarrow {\chi_{\text{c}}}\mathbf{\Theta}_{\text{c}}+(1-\chi_{\text{c}})\mathbf{\Theta}_{\text{c}}^-,\label{softupdate}
\end{equation}
where $\chi_{\text{a}}\,{\in}\,[0,1]$ and $\chi_{\text{c}}\,{\in}\,[0,1]$ are the update factors of the target-actor and the target-critic, respectively. To be specific, the soft-update enables the target-actor and the target-critic to maintain a certain continuity during the update process, so as to better adapt to the continuous control task.

\subsection{Detailed Operations of DeepLSC}
Thus far, we have presented the main elements of the DeepLSC scheme, and Algorithm \ref{alg: algorithm} provides the corresponding pseudocode.
Overall, the operations of DeepLSC include: (i) decide the temporary joint beamforming and trajectory decision ${\pi}_{\text{a}} \left( \mathbf{S}(t); \mathbf{\Theta}_{\text{a}} \right)$ via the eval-actor with the current state $\mathbf{S}(t)$ (lines 7$\sim$8); (ii) decide $\mathbf{a}_{\text{c}}(t)$, $\mathbf{a}_{\text{s}}(t)$, and $\mathbf{a}_{\text{u}}(t)$ via \eqref{actionselection1} (lines 9$\sim$10); (iii) determine the communication beamforming $\mathbf{W}_{\text{c}}(t)$ and the sensing beamforming $\mathbf{W}_{\text{s}}(t)$ via \eqref{actionselection2}, and the movement directions $\mathbf{a}_{\text{u}}(t)$ via \eqref{actionselection3} (lines 11$\sim$13); (iv) obtain the next state $\mathbf{S}(t+1)$ according to \eqref{state1} (line 14); (v) calculate the reward $\{r(t+1)|t=1,\,2,\,\cdots,\,T\}$ via \eqref{rewardfunction1} at the end of each episode (line 16); (vi) form the experience set $\{e(t)=(\mathbf{S}(t),$ $\mathbf{a}(t),$ $r(t+1),$ $\mathbf{S}(t+1))|t=1,\,2,\,\cdots,\,T\}$ and store it in the experience buffer $\mathcal{D}$ (line 17); (vii) randomly sample $N_{\text{e}}$ experience sets from $\mathcal{D}$ and augment each of them with $\Lambda(\omega)$ via \eqref{NewExperience} (lines 18$\sim$20); (viii) integrate the sampled and augmented experience sets to form the mini-batch $\mathcal{B}$ for training the eval-actor and the eval-critic via \eqref{gradientdescent1} and \eqref{gradientdescent2}, respectively (lines 21$\sim$23); and (ix) update the parameters of both the target-actor and the target-critic via \eqref{softupdate} (line 14).

\begin{algorithm}[t]\caption{DeepLSC Scheme}\label{alg: algorithm}
	\begin{algorithmic}[1]
        \State Initialize $N$, $M$, $T$, $\mathbf{b}$, $\mathbf{u}_m^{\text{I}}$, $\mathbf{u}_m^{\text{F}}$, $\{H_m|m\,{\in}\,\mathcal{M}\}$, $v_m$, $v_{\text{Tar}}$.
		\State Initialize $\mathbf{g}(0)$, $H_{\text{Tar}}(0)$, ${\Gamma}_{\text{min}}$, $\delta_1$, $\delta_2$, $\epsilon$, $\kappa$, ${\alpha _{\text{a}}}$, ${\alpha _{\text{c}}}$, $\Upsilon$, $\Omega$.
        \State Initialize $\omega=0$, $\{\sigma_{i,\text{init}}^2|i={\text{c}},{\text{s}},{\text{u}}\}$, $\zeta$, $\mathcal E$, $D$, $N_{\text{e}}$, $\chi_{\text{a}}$, $\chi_{\text{c}}$.
        \State Initialize $\mathbf{\Theta}_{\text{a}}$, $\mathbf{\Theta}_{\text{a}}^-$, $\mathbf{\Theta}_{\text{c}}$, $\mathbf{\Theta}_{\text{c}}^-$.
		\For{$ \text{episode}=1,2, \cdots $}
        \For{$ t=1,2, \cdots, T$}
		\State Input $\mathbf{S}(t)$ into the eval-actor;
        \State Make the temporary decision ${\pi}_{\text{a}} \left( \mathbf{S}(t); \mathbf{\Theta}_{\text{a}} \right)$;
		\State Generate $\mathbf{a}_{\text{c}}(t)$, $\mathbf{a}_{\text{s}}(t)$, and $\mathbf{a}_{\text{u}}(t)$ via \eqref{actionselection1};
        \State Update $\sigma_{i}^2(t)$ to $\sigma_{i,\text{init}}^2{\kappa}^t$;
        \State Obtain $\mathbf{W}_{\text{c}}(t)$ and $\mathbf{W}_{\text{s}}(t)$ via \eqref{actionselection2};
        \State Refine $\mathbf{a}_{\text{u}}(t)$ via \eqref{actionselection3};
		\State Take $\mathbf{A}(t)$ to interact with the environment;
		\State Obtain $\mathbf{S}(t+1)$ via \eqref{state1}.
        \EndFor
		\State Compute $\{r(t+1)|t=1,\,2,\,\cdots,\,T\}$ via \eqref{rewardfunction1};
		\State Form $\{e(t)|t=1,\,2,\,\cdots,\,T\}$ to store into $\mathcal{D}$;
		\State Randomly sample $N_{\text{e}}$ experience sets from $\mathcal{D}$;
        \State Augment each experience set with $\Lambda(\omega)$ via \eqref{NewExperience};
        \State Set $\omega\,\leftarrow\,\text{episode}$ and update $\Lambda(\omega)$;
        \State Merge sampled and augmented experiences to form $\mathcal{B}$;
        \State Calculate $L(\mathbf{\Theta}_{\text{c}} )$ via \eqref{loss1};
        \State Train $\mathbf{\Theta}_{\text{a}}$ via \eqref{gradientdescent1} and $\mathbf{\Theta}_{\text{c}}$ via \eqref{gradientdescent2};
		\State Update $\mathbf{\Theta}_{\text{a}}^-$ and $\mathbf{\Theta}_{\text{c}}^-$ via \eqref{softupdate}.
        \EndFor
	\end{algorithmic}
\end{algorithm}

\section{Performance Evaluation}\label{4}
This section evaluates the performance of DeepLSC based on Python 3.6 simulation platform, where the Keras library \cite{keras} builds all DNN models.

\subsection{Parameter Settings}
\subsubsection{System Setups}
We consider a scenario with one GBS, one target, and $M=4$ UAVs, unless stated otherwise. The GBS is located at $(0, 0, 0)$ m, the initial location of the target is $(-60, 100, 70)$ m, the initial locations of UAVs are uniformly chosen from the area $[-150, -80]$ m ${\times}$ $[60, 150]$ m ${\times}$ $80$ m, and the final locations of UAVs are uniformly chosen from the area $[90, 160]$ m ${\times}$ $[50, 160]$ m ${\times}$ $80$ m.
The movement speed of the target and UAVs are $10$ m per time slot, and the number of time slots within an episode i.e., $T$, is 40 time slots, by default. The movement azimuth and elevation of the target are initialized to $30^{\circ}$, the time correlation coefficients $\mu_{\text{a}}$ and $\mu_{\text{e}}$ for the target movement are $0.9$, and the corresponding asymptotic means and standard deviations, i.e., ${\xi_{\text{a}}}$, ${\xi_{\text{e}}}$, $\sigma_{\phi}$, and $\sigma_{\varphi}$, are $10^{\circ}$.
The number of antennas and the transmit power are $N=6$ and $P_{\text{max}}=40$ dBm, respectively. The noise power ${\sigma}_{\text{b}}^2$ and ${\sigma}_m^2$ are $-80$ dBm, The path loss exponent is 3.2, the reference distance $D_0$ is 1 m, and the path loss $L_0$ for $D_0$ is $-30$ dB.
The sensing SNR requirement $\Gamma_{\text{min}}$ is set to $1$ dB.

\subsubsection{Algorithm Setups}
The GRU and FC layers of DeepLSC contain $128$ neurons, and TABLE \ref{tab:table1} summarizes the detailed hyper-parameters. In the simulation, the following schemes are considered:
\begin{itemize}
    \item DeepLSC: This is our designed DRL-based ISAC scheme for supporting LAE.
    \item DeepLSC-CNE: This scheme replaces the constrained noise-exploration policy with the conventional noise-exploration policy \cite{lillicrap2015continuous}. To meet the signal power constraint for communication and sensing, the scaling factor $\epsilon$ is retained in the conventional noise-exploration policy.
    \item DeepLSC-CER: This scheme replaces the proposed hierarchical experience replay mechanism with the conventional experience replay mechanism \cite{lin1992self}.
    \item DeepLSC-w: This scheme removes the proposed symmetric experience augmentation mechanism to evaluate its effectiveness in DeepLSC.
    \item AC2: This scheme utilizes the actor-critic algorithm \cite{sutton2018reinforcement} with the constrained noise-exploration policy to optimize GBS's beamforming and UAVs' trajectories. Its basic elements, including state, action, and reward function, are exactly the same as DeepLSC. Furthermore, AC2 usually uses a single instantaneous experience set to train its DNN, instead of random historical experience sets.
\end{itemize}

\begin{table}[]
    \centering
    \caption{Algorithm hyper-parameters.}
    \begin{tabular}{|c|c|c|c|}\hline
     \textbf{Hyper-parameter} & \textbf{Value} & \textbf{Hyper-parameter} & \textbf{Value}\\\hline
     Learning rate (${\alpha _{\text{a}}}$) & 1e-4 & Reward coefficient ($\delta_1$/$\delta_2$) & 20/10\\\hline
     Learning rate ($\alpha_{\text{c}}$)  & 2e-4 & Update coefficient ($\chi_{\text{a}}$) & 0.0001\\\hline
     Buffer size ($D$) & 2000 & Update coefficient ($\chi_{\text{c}}$) & 0.0001\\\hline
     Mini-batch size ($N_{\text{e}}$) & 64 & Initial exploration ($ \epsilon _{\text{init}}$) & 0.9\\\hline
     Decaying rate ($\zeta$) & 0.999 & Decaying rate ($\kappa$) & 0.999\\\hline
    \end{tabular}
    \label{tab:table1}
\end{table}

\subsubsection{Metric Setups}
We consider 5000 episodes, each of which contains $T$ time slots in which all UAVs fulfill the preset flight missions. The performance metrics include the communication sum-rate, the sensing SNR, and whether the flight missions of all UAVs are met. Specifically, the communication sum-rate at each episode is a short-term average, which is calculated by averaging $\sum_{t=1}^T{\text{R}_{\text{total}}}(t)$ over the previous 200 episodes; and the short-term average sensing SNR at each episode is calculated by averaging $\sum_{t=1}^T{\text{SNR}_{\text{Tar}}}(t)/T$ over the previous 200 episodes. To ensure the reliability of experiments, all simulations run 20 times to obtain the average result.

\subsection{Learning Curves}\label{LearningCurves}
\begin{figure}[t]
	\centering
	\includegraphics[height=6.36cm]{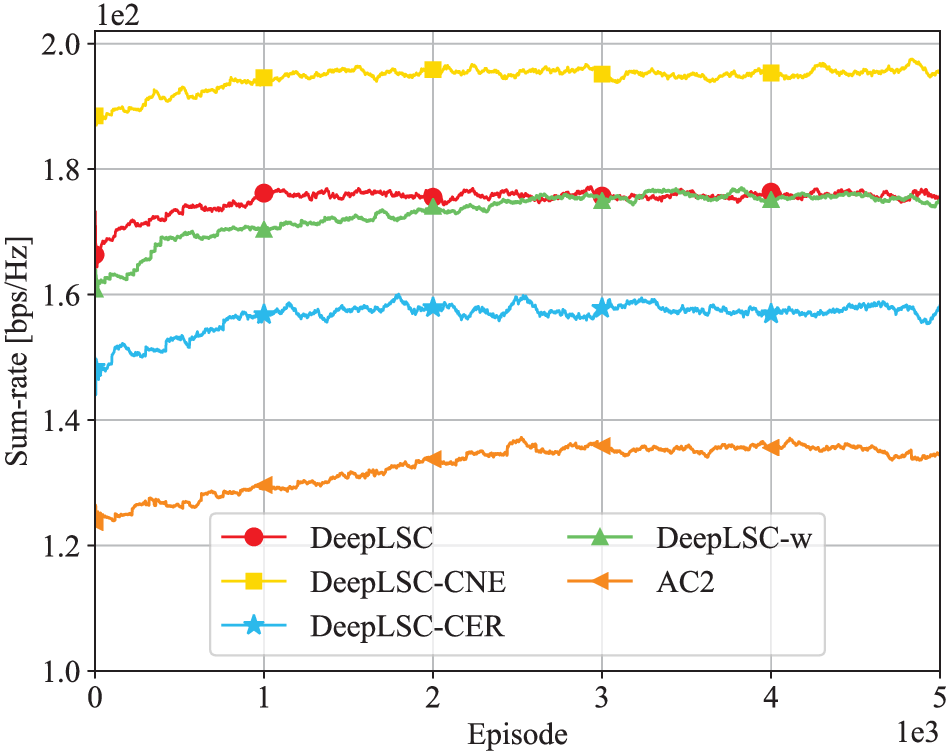}
	\caption{Communication sum-rate achieved by various schemes.}
	\label{Sumrate1}
\end{figure}

\begin{figure}[t]
	\centering
	\includegraphics[height=6.36cm]{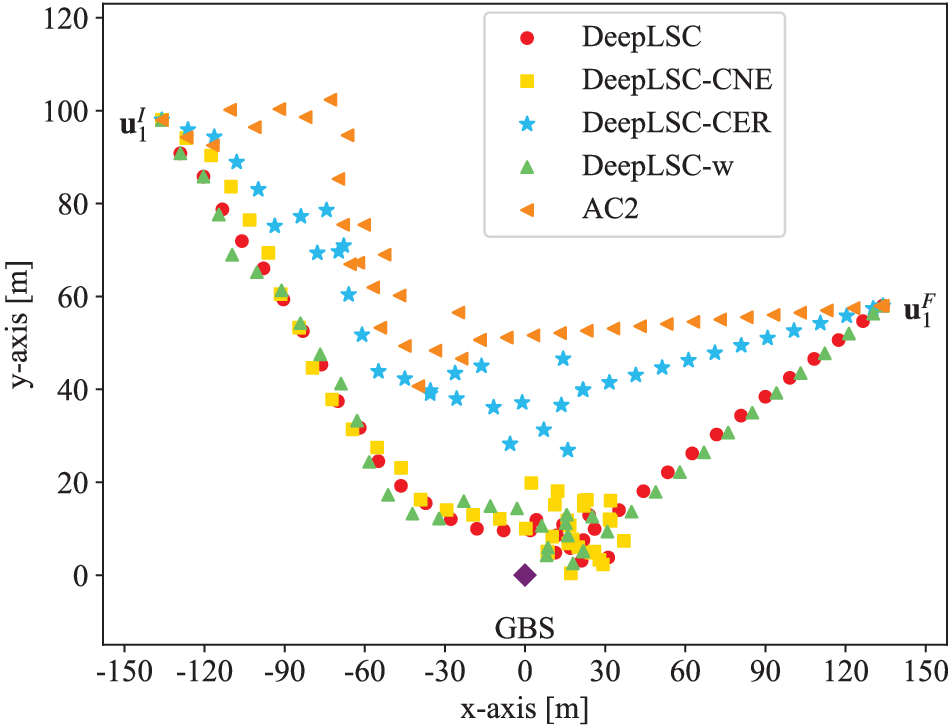}
	\caption{Flight trajectory of a specific UAV under various schemes.}
	\label{Trajectory1}
\end{figure}

Fig. \ref{Sumrate1} and TABLE \ref{tab:table2} provide the communication sum-rate and average sensing SNR achieved by various schemes, respectively. As can be seen from Fig. \ref{Sumrate1}, due to the random mobility of the target, all the curves are in fluctuation. In other words, in different episodes, the joint beamforming and trajectory strategies derived from the same scheme may be different. Fig. \ref{Trajectory1} depicts the flight trajectory of a specific UAV under various schemes in the last simulation episode.

\begin{table}[]
    \centering
    \caption{Average sensing SNR [dB] of various schemes.}
    \begin{tabular}{|c|c|c|}\hline
     \textbf{Scheme} & \textbf{Average sensing SNR} & \textbf{Target sensing SNR}\\\hline
     DeepLSC & 1.12 & 1.00\\\hline
     DeepLSC-CNE & 1.24 & 1.00\\\hline
     DeepLSC-CER & 1.03 & 1.00\\\hline
     DeepLSC-w & 1.12 & 1.00\\\hline
     AC2 & 0.86 & 1.00\\\hline
    \end{tabular}
    \label{tab:table2}
\end{table}

\textit{1) DeepLSC Versus DeepLSC-CNE:} As presented in Fig. \ref{Sumrate1}, among all the schemes, DeepLSC-CNE achieves the highest communication sum-rate while guaranteeing the given sensing SNR constraint. To be specific, compared with DeepLSC, the communication sum-rate gain of DeepLSC-CNE is around 11.14\%. This is because DeepLSC-CNE is designed to maximize the communication sum-rate and satisfy the average sensing SNR requirement, while ignoring the flight mission constraint. As a result, under DeepLSC-CNE, UAVs tend to stay in a specific area instead of flying from initial locations to desired locations within each episode, as shown in Fig. \ref{Trajectory1}. On the contrary, owing to the constrained noise-exploration policy, DeepLSC can well guide UAVs to reach their desired locations within an episode, although sacrificing both communication and sensing performance to some certain. This result demonstrates the effectiveness of the proposed constrained noise-exploration policy in fulfilling the flight mission.

\textit{2) DeepLSC Versus AC2:} Similar to DeepLSC, AC2 employs the constrained noise-exploration policy to decide the movement directions of UAVs. It, however, not only fails to satisfy the given average sensing SNR requirement, but also achieves a much lower communication sum-rate than DeepLSC. In particular, compared with DeepLSC, the sum-rate reduction of AC2 is more than 22.73\%. The underlying reasons are as follows. First, before executing the constrained noise-exploration policy, DeepLSC derives a deterministic joint beamforming and trajectory decision, whereas AC2 usually generates a probability distribution corresponding to all possible actions and randomly samples an action from this distribution, resulting in inefficient decisions. Second, during DNN training, DeepLSC makes full use of random historical experience sets, but AC2 relies on a single instantaneous experience set. As a consequence, the parameters of the DNN are strongly correlated before and after the update, which reduces the efficiency of the decision-making \cite{sutton2018reinforcement}. Besides, in DeepLSC, the target-actor and target-critic are utilized to assist in training eval-actor and eval-critic, which further boosts the stability of the algorithm, whilst AC2 lacks this consideration.

\textit{3) DeepLSC Versus DeepLSC-CER:} Unlike AC2, the underpinning algorithm in DeepLSC-CER is DDPG \cite{mnih2015human}, which circumvents the above issues well. Thus, in comparison to AC2, while meeting the preset average sensing SNR requirement, DeepLSC-CER yields a significant communication sum-rate gain of about 15.88\%. DeepLSC-CER, however, does not cater well to the formulated optimization problem. Specifically, the formulated optimization problem is an episode task, in the sense that all experiences generated within an episode should be jointly used to train the DNN. Unfortunately, the DNN training mechanism adopted in DeepLSC-CER, i.e., conventional experience replay \cite{lin1992self}, separately collects and utilizes the experiences from different time slots. The consequence is that all experiences generated within an episode cannot be guaranteed to appear in full during training. By contrast, benefiting from the hierarchical experience replay mechanism, DeepLSC is able to take advantage of experience sets, each composed of all experiences generated within an episode. Thus, DeepLSC can well handle the formulated problem and further improve the sum-rate of LAE-oriented ISAC systems: compared with DeepLSC-CER, the improvement attained by DeepLSC is about 11.68\%.

\textit{4) DeepLSC Versus DeepLSC-w:} From Fig. \ref{Sumrate1}, it can also be found that DeepLSC-w experiences more episodes before convergence compared to DeepLSC, DeepLSC-CNE, and DeepLSC-CER, although they are all online learning schemes. In particular, compared with the other three schemes, DeepLSC-w consumes over 130.60\% more episodes to converge. The main reasons are illustrated below. In DeepLSC-w, the experience sets can only be obtained through interactions with the environment. In general, to efficiently train the DNN, a sufficient number of experience sets are essential. As a result, the learning process of the agent is slow. Unlike DeepLSC-w, based on the limited existing experience sets, DeepLSC, DeepLSC-CNE, and DeepLSC-CER schemes exploit the symmetric experience augmentation mechanism to enrich the available experience sets. Therefore, these three schemes can acquire massive experience sets in a short time, thereby significantly accelerating the convergence speed. On the other hand, in terms of communication and sensing performance, DeepLSC-w is the same as DeepLSC after convergence; and the UAV trajectories generated by the two are also similar, as provided in Fig. \ref{Trajectory1}. This is because the proposed symmetric experience augmentation mechanism will not impair algorithm performance despite achieving faster convergence.

\begin{figure}[t]
	\centering
	\includegraphics[height=6.36cm]{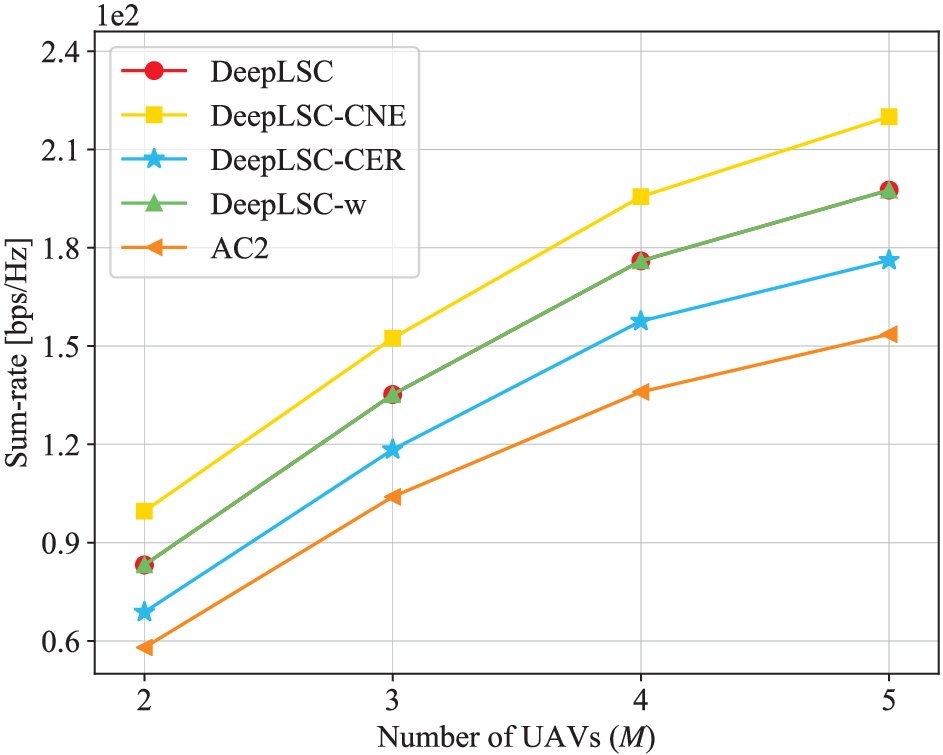}
	\caption{Sum-rate of various schemes under different numbers of UAVs.}
	\label{Sumrate2}
\end{figure}

\subsection{Robustness to Different Numbers of UAVs}\label{RobustnessUAVs}
This subsection evaluates the robustness of various schemes against different numbers of UAVs, in which $M$ increases from 2 to 5 with a step size of 1. Fig. \ref{Sumrate2} and TABLE \ref{tab:table3} present the communication sum-rate and constraint satisfactory of various schemes, respectively.

It can be observed that as $M$ increases, the communication sum-rate achieved by all schemes are gradually improved. This is because when $M$ increases, the GBS can support more parallel transmissions. At this time, under a reasonable beamforming design, spatial multiplexing can be further improved, thereby increasing the communication sum-rate. In addition, with the increase of $M$, the sum-rate gap between DeepLSC and DeepLSC-CER/AC2 increases. For example, when $M$ increases from 2 to 5, the sum-rate gain of DeepLSC increases from 14.43 bps/Hz to 21.42 bps/Hz compared with DeepLSC-CER.
On the other hand, as summarized in TABLE \ref{tab:table3}, due to the inefficient DNN training mechanism, both DeepLSC-CER and AC2 cannot always meet the average sensing SNR requirement under different values of $M$, whereas DeepLSC-CNE fails to fulfill the flight mission since it lacks the guidance of the constrained noise-exploration policy. Thus, among all the schemes, only DeepLSC and DeepLSC-w satisfy all constraints under various settings. DeepLSC-w, however, converges much slower than DeepLSC, as elaborated in Section \ref{LearningCurves}. Overall, the above observations exemplify that compared with other schemes, DeepLSC is more robust against different numbers of UAVs in LAE-oriented ISAC systems.

\begin{table}[]
    \centering
    \caption{Performance of various schemes in satisfying the average sensing SNR and flight mission constraints under different numbers of UAVs.}
    \begin{tabular}{|c|c|c|c|c|}\hline
    \multicolumn{5}{|c|}{\textbf{Average sensing SNR constraint}}\\\hline
     & $M=2$ & $M=3$ & $M=4$ & $M=5$\\\hline
     DeepLSC & \checkmark & \checkmark & \checkmark & \checkmark\\\hline
     DeepLSC-CNE & \checkmark & \checkmark & \checkmark & \checkmark\\\hline
     DeepLSC-CER & \checkmark & \checkmark & \checkmark & $\times$\\\hline
     DeepLSC-w & \checkmark & \checkmark & \checkmark & \checkmark\\\hline
     AC2 & \checkmark & \checkmark & $\times$ & $\times$\\\hline
     \multicolumn{5}{|c|}{\textbf{Flight mission constraint}}\\\hline
     & $M=2$ & $M=3$ & $M=4$ & $M=5$\\\hline
     DeepLSC & \checkmark & \checkmark & \checkmark & \checkmark\\\hline
     DeepLSC-CNE & $\times$ & $\times$ & $\times$ & $\times$\\\hline
     DeepLSC-CER & \checkmark & \checkmark & \checkmark & \checkmark\\\hline
     DeepLSC-w & \checkmark & \checkmark & \checkmark & \checkmark\\\hline
     AC2 & \checkmark & \checkmark & \checkmark & \checkmark\\\hline
    \end{tabular}
    \label{tab:table3}
\end{table}

\subsection{Robustness to Different Numbers of Time Slots Within a Flight Period}\label{RobustnessFlightPeriods}
\begin{figure}[t]
	\centering
	\includegraphics[height=6.36cm]{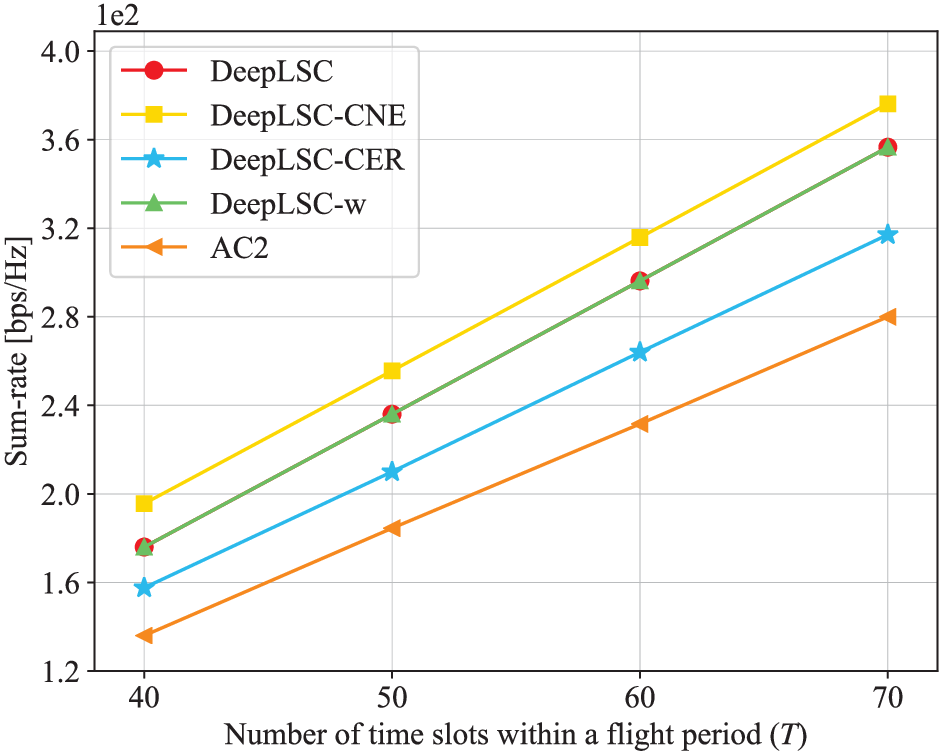}
	\caption{Sum-rate of various schemes under different numbers of time slots within a flight period.}
	\label{Sumrate3}
\end{figure}

This simulation evaluates the robustness of various schemes to different numbers of time slots within a flight period, in which $T$ increases from 40 to 70 with a step size of 10. The communication sum-rate and constraint satisfactory of various schemes are shown in Fig. \ref{Sumrate3} and TABLE \ref{tab:table4}, respectively.

As expected, the communication sum-rate attained by various schemes improves monotonically as $T$ increases. The main reasons are below. As $T$ increases, the number of time slots during which all UAVs can fly freely without the flight mission constraint increases. In this regard, all schemes possess more time degrees of freedom to optimize the joint beamforming and trajectory strategy, so as to maximize the communication sum-rate while satisfying the average sensing SNR requirement. Besides, under different $T$, DeepLSC, DeepLSC-CER, and DeepLSC-w schemes can always satisfy all preset constraints, whilst both DeepLSC-CNE and AC2 fail. Thanks to a more intelligent and efficient strategy, DeepLSC achieves a higher communication sum-rate than DeepLSC-CER. To be specific, compared with DeepLSC-CER, DeepLSC yields a more than 11.68\% sum-rate gain for all simulation setups.

More importantly, as $T$ increases, the communication sum-rate gap between DeepLSC and DeepLSC-CNE keeps almost unchanged, whereas that between DeepLSC and DeepLSC-CER/AC2 increases. In particular, when $M$ increases from 40 to 70, the sum-rate improvement of DeepLSC increases (i) from 18.42 bps/Hz to 39.56 bps/Hz compared with DeepLSC-CER and (ii) from 40.03 bps/Hz to 76.63 bps/Hz compared with AC2. The above results show that under different numbers of time slots within a flight period, DeepLSC is more robust than other schemes.


\begin{table}[]
    \centering
    \caption{Performance of various schemes in satisfying the average sensing SNR and flight mission constraints under different numbers of time slots within a flight period.}
    \begin{tabular}{|c|c|c|c|c|}\hline
    \multicolumn{5}{|c|}{\textbf{Average sensing SNR constraint}}\\\hline
     & $T=40$ & $T=50$ & $T=60$ & $T=70$\\\hline
     DeepLSC & \checkmark & \checkmark & \checkmark & \checkmark\\\hline
     DeepLSC-CNE & \checkmark & \checkmark & \checkmark & \checkmark\\\hline
     DeepLSC-CER & \checkmark & \checkmark & \checkmark & \checkmark\\\hline
     DeepLSC-w & \checkmark & \checkmark & \checkmark & \checkmark\\\hline
     AC2 & $\times$ & $\times$ & $\times$ & $\times$\\\hline
     \multicolumn{5}{|c|}{\textbf{Flight mission constraint}}\\\hline
     & $T=40$ & $T=50$ & $T=60$ & $T=70$\\\hline
     DeepLSC & \checkmark & \checkmark & \checkmark & \checkmark\\\hline
     DeepLSC-CNE & $\times$ & $\times$ & $\times$ & $\times$\\\hline
     DeepLSC-CER & \checkmark & \checkmark & \checkmark & \checkmark\\\hline
     DeepLSC-w & \checkmark & \checkmark & \checkmark & \checkmark\\\hline
     AC2 & \checkmark & \checkmark & \checkmark & \checkmark\\\hline
    \end{tabular}
    \label{tab:table4}
\end{table}

\section{Conclusion}\label{5}
This paper put forth a new ISAC scheme for LAE, termed DeepLSC, based on the DRL technique. By jointly optimizing GBS's beamforming and UAVs' trajectories, DeepLSC aims to maximize the expected communication sum-rate over the flight period, subject to the average sensing SNR requirement, flight mission, collision avoidance, and maximum transmit power constraints. DeepLSC, on the one hand, is model-free since it does not require prior mobility information of the target; on the other hand, it is capable of episode tasks thanks to an appropriate structural design. First, to meet various constraints, a constrained noise-exploration policy and a reward function were judiciously designed in DeepLSC for action selection and action evaluation, respectively. Thereafter, we developed a hierarchical experience replay mechanism for DNN training, where all experiences within an episode are jointly utilized to enable DeepLSC to efficiently learn from episode tasks. Besides, to promote the convergence speed of DeepLSC, a symmetric experience augmentation mechanism was further proposed. Simulation results demonstrated compared with other schemes, DeepLSC (i) yields a much higher communication sum-rate while meeting all constraints, (ii) converges faster, and (iii) is more robust against different settings.

\bibliography{reference}

\begin{thebibliography}{10}
\providecommand{\url}[1]{#1}
\csname url@samestyle\endcsname
\providecommand{\newblock}{\relax}
\providecommand{\bibinfo}[2]{#2}
\providecommand{\BIBentrySTDinterwordspacing}{\spaceskip=0pt\relax}
\providecommand{\BIBentryALTinterwordstretchfactor}{4}
\providecommand{\BIBentryALTinterwordspacing}{\spaceskip=\fontdimen2\font plus
\BIBentryALTinterwordstretchfactor\fontdimen3\font minus \fontdimen4\font\relax}
\providecommand{\BIBforeignlanguage}[2]{{%
\expandafter\ifx\csname l@#1\endcsname\relax
\typeout{** WARNING: IEEEtran.bst: No hyphenation pattern has been}%
\typeout{** loaded for the language `#1'. Using the pattern for}%
\typeout{** the default language instead.}%
\else
\language=\csname l@#1\endcsname
\fi
#2}}
\providecommand{\BIBdecl}{\relax}
\BIBdecl

\bibitem{jiang20236g}
Y.~Jiang, X.~Li, G.~Zhu, H.~Li, J.~Deng, and Q.~Shi, ``{6G} non-terrestrial networks enabled low-altitude economy: {Opportunities} and challenges,'' \emph{arXiv preprint arXiv:2311.09047}, 2023.

\bibitem{LAESurvey}
\BIBentryALTinterwordspacing
{China Telecom, Ericsson, Nokia, Huawei, ZTE, CICT Mobile, OPPO, Xiaomi, vivo, Lenovo, Qualcomm, Mediatek, UNISOC}, ``The low-altitude network by integrated sensing and communication,'' White Paper, Feb. 2024. [Online]. Available: \url{https://www.zte.com.cn/content/dam/zte-site/res-www-zte-com-cn/mediares/zte/%E6%97%A0%E7%BA%BF%E6%8E%A5%E5%85%A5/%E7%99%BD%E7%9A%AE%E4%B9%A6/Low_altitude_network_by_ISAC.pdf}
\BIBentrySTDinterwordspacing

\bibitem{mu2023uav}
J.~Mu, R.~Zhang, Y.~Cui, N.~Gao, and X.~Jing, ``{UAV} meets integrated sensing and communication: {Challenges} and future directions,'' \emph{IEEE Commun. Mag.}, vol.~61, no.~5, pp. 62--67, Jan. 2023.

\bibitem{liu2022integrated}
F.~Liu, Y.~Cui, C.~Masouros, J.~Xu, T.~X. Han, Y.~C. Eldar, and S.~Buzzi, ``Integrated sensing and communications: {Toward} dual-functional wireless networks for {6G} and beyond,'' \emph{IEEE J. Sel. Areas in Commun.}, vol.~40, no.~6, pp. 1728--1767, Mar. 2022.

\bibitem{zhang2021uav}
K.~Zhang and C.~Shen, ``{UAV} aided integrated sensing and communications,'' in \emph{Proc. IEEE VTC}, Norman, OK, USA, Sept. 2021, pp. 1--6.

\bibitem{zhang2020age}
S.~Zhang, H.~Zhang, Z.~Han, H.~V. Poor, and L.~Song, ``Age of information in a cellular internet of {UAVs}: {Sensing} and communication trade-off design,'' \emph{IEEE Trans. Wireless Commun.}, vol.~19, no.~10, pp. 6578--6592, Jun. 2020.

\bibitem{wang2020constrained}
X.~Wang, Z.~Fei, J.~A. Zhang, J.~Huang, and J.~Yuan, ``Constrained utility maximization in dual-functional radar-communication {multi-UAV} networks,'' \emph{IEEE Trans. Commun.}, vol.~69, no.~4, pp. 2660--2672, Dec. 2020.

\bibitem{chang2022integrated}
B.~Chang, W.~Tang, X.~Yan, X.~Tong, and Z.~Chen, ``Integrated scheduling of sensing, communication, and control for {mmWave/THz} communications in cellular connected {UAV} networks,'' \emph{IEEE J. Sel. Areas in Commun.}, vol.~40, no.~7, pp. 2103--2113, Mar. 2022.

\bibitem{ding2023multi}
W.~Ding, C.~Chen, Y.~Fang, L.~Qiu, X.~Li, X.~Wang, and J.~Xu, ``Multi-{UAV}-enabled integrated sensing and communications: {Joint} {UAV} placement and power control,'' in \emph{Proc. IEEE Globecom Workshops}, Kuala Lumpur, Malaysia, Dec. 2023, pp. 842--847.

\bibitem{bayessa2024joint}
G.~A. Bayessa, R.~Chai, C.~Liang, D.~K. Jain, and Q.~Chen, ``Joint {UAV} deployment and precoder optimization for multicasting and target sensing in {UAV}-assisted {ISAC} networks,'' \emph{IEEE Internet Things J.}, vol.~11, no.~20, pp. 33\,392--33\,405, Jul. 2024.

\bibitem{yu2024security}
X.~Yu, J.~Xu, N.~Zhao, X.~Wang, and D.~Niyato, ``Security enhancement of {ISAC} via {IRS-UAV},'' \emph{IEEE Trans. Wireless Commun.}, vol.~23, no.~10, pp. 15\,601--15\,612, Jul. 2024.

\bibitem{wu2023uavs}
J.~Wu, W.~Yuan, and L.~Hanzo, ``When {UAVs} meet {ISAC}: {Real-time} trajectory design for secure communications,'' \emph{IEEE Trans. Veh. Technol.}, vol.~72, no.~12, pp. 16\,766--16\,771, Dec. 2023.

\bibitem{hu2022trajectory}
S.~Hu, X.~Yuan, W.~Ni, and X.~Wang, ``Trajectory planning of cellular-connected {UAV} for communication-assisted radar sensing,'' \emph{IEEE Trans. Commun.}, vol.~70, no.~9, pp. 6385--6396, Aug. 2022.

\bibitem{liu2023sensing}
Y.~Liu, S.~Liu, X.~Liu, Z.~Liu, and T.~S. Durrani, ``Sensing fairness-based energy efficiency optimization for {UAV} enabled integrated sensing and communication,'' \emph{IEEE Wireless Commun. Lett.}, vol.~12, no.~10, pp. 1702--1706, Jun. 2023.

\bibitem{liu2024uav}
Z.~Liu, X.~Liu, Y.~Liu, V.~C. Leung, and T.~S. Durrani, ``{UAV} assisted integrated sensing and communications for internet of things: {3D} trajectory optimization and resource allocation,'' \emph{IEEE Trans. Wireless Commun.}, vol.~23, no.~8, pp. 8654--8667, Jan. 2024.

\bibitem{liu2024secure}
Y.~Liu, X.~Liu, Z.~Liu, Y.~Yu, M.~Jia, Z.~Na, and T.~S. Durrani, ``Secure rate maximization for isac-uav assisted communication amidst multiple eavesdroppers,'' \emph{IEEE Trans. Veh. Technol.}, vol.~73, no.~10, pp. 15\,843--15\,847, Jun. 2024.

\bibitem{zhang2024secure}
J.~Zhang, J.~Xu, W.~Lu, N.~Zhao, X.~Wang, and D.~Niyato, ``Secure transmission for {IRS}-aided {UAV-ISAC} networks,'' \emph{IEEE Trans. Wireless Commun.}, vol.~23, no.~9, pp. 12\,256--12\,269, Apr. 2024.

\bibitem{jing2024isac}
X.~Jing, F.~Liu, C.~Masouros, and Y.~Zeng, ``{ISAC} from the sky: {UAV} trajectory design for joint communication and target localization,'' \emph{IEEE Trans. Wireless Commun.}, vol.~23, no.~10, pp. 12\,857--12\,872, May 2024.

\bibitem{meng2022throughput}
K.~Meng, Q.~Wu, S.~Ma, W.~Chen, K.~Wang, and J.~Li, ``Throughput maximization for {UAV}-enabled integrated periodic sensing and communication,'' \emph{IEEE Trans. Wireless Commun.}, vol.~22, no.~1, pp. 671--687, Aug. 2022.

\bibitem{meng2022uav}
K.~Meng, Q.~Wu, S.~Ma, W.~Chen, and T.~Q. Quek, ``{UAV} trajectory and beamforming optimization for integrated periodic sensing and communication,'' \emph{IEEE Wireless Commun. Lett.}, vol.~11, no.~6, pp. 1211--1215, Mar. 2022.

\bibitem{lyu2022joint}
Z.~Lyu, G.~Zhu, and J.~Xu, ``Joint maneuver and beamforming design for {UAV}-enabled integrated sensing and communication,'' \emph{IEEE Trans. Wireless Commun.}, vol.~22, no.~4, pp. 2424--2440, Oct. 2022.

\bibitem{wu2024joint}
Z.~Wu, X.~Li, Y.~Cai, and W.~Yuan, ``Joint trajectory and resource allocation design for {RIS}-assisted {UAV}-enabled {ISAC} systems,'' \emph{IEEE Wireless Commun. Lett.}, vol.~13, no.~5, pp. 1384--1388, Feb. 2024.

\bibitem{deng2023beamforming}
C.~Deng, X.~Fang, and X.~Wang, ``Beamforming design and trajectory optimization for {UAV}-empowered adaptable integrated sensing and communication,'' \emph{IEEE Trans. Wireless Commun.}, vol.~22, no.~11, pp. 8512--8526, Apr. 2023.

\bibitem{van2024joint}
T.~Van~Chien, M.~D. Cong, N.~C. Luong, T.~N. Do, D.~I. Kim, and S.~Chatzinotas, ``Joint computation offloading and target tracking in integrated sensing and communication enabled {UAV} networks,'' \emph{IEEE Commun. Lett.}, vol.~28, no.~6, pp. 1327--1331, Apr. 2024.

\bibitem{wu2023interplay}
J.~Wu, W.~Yuan, and L.~Bai, ``On the interplay between sensing and communications for {UAV} trajectory design,'' \emph{IEEE Internet Things J.}, vol.~10, no.~23, pp. 20\,383--20\,395, Jun. 2023.

\bibitem{zhang2024joint}
R.~Zhang, Y.~Zhang, R.~Tang, H.~Zhao, Q.~Xiao, and C.~Wang, ``A joint {UAV} trajectory, user association, and beamforming design strategy for multi-{UAV} assisted {ISAC} systems,'' \emph{IEEE Internet Things J.}, vol.~11, no.~8, pp. 29\,360--29\,374, Jul. 2024.

\bibitem{zhang2022trajectory}
T.~Zhang, K.~Zhu, S.~Zheng, D.~Niyato, and N.~C. Luong, ``Trajectory design and power control for joint radar and communication enabled {multi-UAV} cooperative detection systems,'' \emph{IEEE Trans. Commun.}, vol.~71, no.~1, pp. 158--172, Nov. 2022.

\bibitem{xie2024distributed}
Z.~Xie, Z.~Wang, Z.~Zhang, J.~Wang, Z.~Jiang, and Z.~Han, ``Distributed {UAV} swarm for device-free integrated sensing and communication relying on multi-agent reinforcement learning,'' \emph{IEEE Trans. Veh. Technol.}, Aug. 2024, early access, doi: 10.1109/TVT.2024.3438854.

\bibitem{chen2024drl}
X.~Chen, X.~Cao, L.~Xie, and Y.~He, ``{DRL}-based joint trajectory planning and beamforming optimization in aerial {RIS}-assisted {ISAC} system,'' in \emph{Proc. IEEE iWRF\&AT}, Shenzhen, China, Jul. 2024, pp. 510--515.

\bibitem{cho2024enhancing}
H.~Cho, S.~Yoo, B.~C. Jung, and J.~Kang, ``Enhancing battlefield awareness: {An} aerial {RIS}-assisted {ISAC} system with deep reinforcement learning,'' \emph{arXiv preprint arXiv:2405.20168}, 2024.

\bibitem{fontanesi2024deep}
G.~Fontanesi, A.~Guerra, F.~Guidi, J.~A. V{\'a}squez-Peralvo, N.~Shlezinger, A.~Zanella, E.~Lagunas, S.~Chatzinotas, D.~Dardari, and P.~M. Djuri{\'c}, ``A deep-{NN} beamforming approach for dual function radar-communication {THz} {UAV},'' \emph{arXiv preprint arXiv:2405.17015}, 2024.

\bibitem{qin2023deep}
Y.~Qin, Z.~Zhang, X.~Li, W.~Huangfu, and H.~Zhang, ``Deep reinforcement learning based resource allocation and trajectory planning in integrated sensing and communications {UAV} network,'' \emph{IEEE Trans. Wireless Commun.}, vol.~22, no.~11, pp. 8158--8169, Mar. 2023.

\bibitem{dai2024joint}
C.~Dai, T.~Wu, G.~Sun, Y.~Zuo, Z.~Guo, and F.~Xiao, ``Joint {UAV} trajectory and beamforming design for ris-aided integrated sensing and communication system,'' in \emph{Proc. IEEE INFOCOM Workshops}, Vancouver, BC, Canada, May 2024, pp. 1--6.

\bibitem{cui2023specific}
Y.~Cui, Q.~Zhang, Z.~Feng, F.~Liu, C.~Shi, J.~Fan, and P.~Zhang, ``Specific beamforming for {multi-UAV} networks: A dual identity-based {ISAC} approach,'' in \emph{Proc. IEEE ICC}, Rome, Italy, May 2023, pp. 4979--4985.

\bibitem{cheng2024networked}
G.~Cheng, X.~Song, Z.~Lyu, and J.~Xu, ``Networked {ISAC} for low-altitude economy: Coordinated transmit beamforming and {UAV} trajectory design,'' \emph{arXiv preprint arXiv:2406.16946}, 2024.

\bibitem{sutton2018reinforcement}
R.~S. Sutton and A.~G. Barto, \emph{Reinforcement learning: An introduction}.\hskip 1em plus 0.5em minus 0.4em\relax Cambridge, MA, USA:MIT press, 2018.

\bibitem{mnih2015human}
V.~Mnih, K.~Kavukcuoglu, D.~Silver, A.~A. Rusu, J.~Veness, M.~G. Bellemare, A.~Graves, M.~Riedmiller, A.~K. Fidjeland, G.~Ostrovski \emph{et~al.}, ``Human-level control through deep reinforcement learning,'' \emph{nature}, vol. 518, no. 7540, pp. 529--533, Feb. 2015.

\bibitem{ye2021multi}
{X. Ye, Y. Yu, and L. Fu}, ``Multi-channel opportunistic access for heterogeneous networks based on deep reinforcement learning,'' \emph{IEEE Trans. Wireless Commun.}, vol.~21, no.~2, pp. 794--807, Feb. 2022.

\bibitem{lillicrap2015continuous}
T.~Lillicrap, J.~Hunt, A.~Pritzel, N.~Heess, T.~Erez, Y.~Tassa, D.~Silver, and D.~Wierstra, ``Continuous control with deep reinforcement learning,'' in \emph{Proc. Int. Conf. Learn. Represent.}, San Juan, Puerto Rico, May 2016, pp. 1--14.

\bibitem{lin1992self}
L.-J. Lin, ``Self-improving reactive agents based on reinforcement learning, planning and teaching,'' \emph{Mach. Learn.}, vol.~8, pp. 293--321, May 1992.

\bibitem{tabassum2019fundamentals}
H.~Tabassum, M.~Salehi, and E.~Hossain, ``Fundamentals of mobility-aware performance characterization of cellular networks: {A} tutorial,'' \emph{IEEE Commun. Surveys Tuts.}, vol.~21, no.~3, pp. 2288--2308, Mar. 2019.

\bibitem{sutton1999policy}
R.~S. Sutton, D.~McAllester, S.~Singh, and Y.~Mansour, ``Policy gradient methods for reinforcement learning with function approximation,'' in \emph{Proc. Adv. Neural Inf. Process. Syst.}, Denver, USA, Nov. 1999, pp. 1057--1063.

\bibitem{seress2003permutation}
{\'A}.~Seress, \emph{Permutation group algorithms}.\hskip 1em plus 0.5em minus 0.4em\relax Cambridge University Press, 2003, no. 152.

\bibitem{keras}
\BIBentryALTinterwordspacing
F.~Chollet, \emph{Keras: {The} python deep learning library}, Jun. 2018. [Online]. Available: \url{https://keras.io.}
\BIBentrySTDinterwordspacing

\end{thebibliography}

\bibliographystyle{IEEEtran}

\end{document}